\documentclass[12pt]{article}
\pdfoutput=1
\usepackage{amsmath,amssymb,color,epsfig,draft}
\usepackage{graphicx}
\usepackage{setspace}
\usepackage{braket}
\usepackage{physics}
\usepackage{slashed}
\usepackage{simplewick}
\usepackage{cancel}
\usepackage{extarrows}
\usepackage[offset=1.25em]{simpler-wick}
\usepackage{tcolorbox}

\usepackage{tikz-cd}

\makeatletter
\@addtoreset{equation}{section}
\makeatother

\newcommand{\bpm}{\begin{pmatrix}}
\newcommand{\epm}{\end{pmatrix}}

\def\0{{\sst{(0)}}}
\def\1{{\sst{(1)}}}
\def\2{{\sst{(2)}}}
\def\3{{\sst{(3)}}}
\def\4{{\sst{(4)}}}
\def\5{{\sst{(5)}}}
\def\6{{\sst{(6)}}}
\def\7{{\sst{(7)}}}
\def\8{{\sst{(8)}}}
\def\sst#1{{\scriptscriptstyle #1}}

\def\cH{{{\cal H}}}

\def\ie{\begin{equation}\begin{aligned}}
\def\fe{\end{aligned}\end{equation}}

\newcommand{\strtwo}[2]{
    \quad\begin{gathered}
        \begin{tikzpicture}
            \draw (0,0) ellipse (1 and 0.5);
            \node[above] at (0,0.5) {$\scriptstyle #1$};
            \node[below] at (0,-0.5) {$\scriptstyle #2$};
        \end{tikzpicture}
    \end{gathered}\quad
}

\newtheorem{conj}{Conjecture}
\newtheorem{defn}{Definition}

\begin{document}

\renewcommand\arraystretch{1.5}

\begin{titlepage}

\begin{center}

\title{Fortuity in the D1-D5 system}

\author{Chi-Ming Chang$^{a,b}$, Ying-Hsuan Lin$^c$, and Haoyu Zhang$^{d,e}$}

\address{${}^a$Yau Mathematical Sciences Center (YMSC), Tsinghua University, Beijing 100084, China}

\address{${}^b$Beijing Institute of Mathematical Sciences and Applications (BIMSA) \\ Beijing 101408, China}

\address{${}^c$Jefferson Physical Laboratory, Harvard University, Cambridge, MA 02138, USA}

\address{${}^d$George P. \&  Cynthia Woods Mitchell Institute for Fundamental Physics and Astronomy,
Texas A\&M University, College Station, TX 77843, USA}

\address{${}^e$ Department of Physics and Astronomy \\
University of Southern California,
Los Angeles, CA 90089, USA }

\email{cmchang@tsinghua.edu.cn, yhlin@alum.mit.edu, zhanghaoyu@tamu.edu }

\end{center}

\vfill

\begin{abstract}

We reformulate the lifting problem in the D1-D5 CFT as a supercharge cohomology problem, and enumerate BPS states according to the fortuitous/monotone classification. Working in the deformed $T^4$ symmetric orbifold theory, we give precise definitions of monotone and fortuitous cohomology classes generalizing the definitions in \cite{Chang:2024zqi} and illustrate them in the $N=1$ theory. For $N=2$, we construct the cohomology explicitly and match it to the exact BPS partition function. We further describe how to assemble BPS states at smaller $N$ into BPS states at larger $N$, and interpret their holographic duals as black hole bound states and massive stringy excitations on smooth horizonless (e.g. Lunin-Mathur) geometries. 

\end{abstract}

\vfill

\end{titlepage}

\tableofcontents

\section{Introduction and summary of results}

Recent progress in identifying the non-graviton states \cite{Chang:2022mjp,Choi:2022caq,Choi:2023znd,Budzik:2023vtr,Choi:2023vdm,deMelloKoch:2024pcs}, also dubbed fortuitous states \cite{Chang:2024zqi}, in the $\frac{1}{16}$-BPS sector of \(\mathcal{N}=4\) supersymmetric Yang-Mills (SYM) has significantly enhanced our understanding of black hole microstates in the framework of the AdS/CFT correspondence \cite{Maldacena:1997re}. This development goes beyond the standard index counting \cite{Kinney:2005ej,Cabo-Bizet:2018ehj,Choi:2018hmj} and examines the supercharge $Q$-cohomology \cite{Kinney:2005ej,Grant:2008sk,Chang:2013fba}, which is isomorphic to the space of BPS states, and offers much-refined information about the microstates. In a nutshell, fortuitous states have a distinctive property that prevents them from remaining BPS at large $N$, and yet they dominate the entropy. Based on these attributes, it was argued and conjectured in \cite{Chang:2024zqi} that fortuitous states are dual to typical black hole microstates, while graviton states are dual to perturbative excitations and smooth horizonless geometries.

In ${\cal N}=4$ SYM, fortuitous states arise due to non-trivial trace relations at finite $N$ among gauge-invariant operators. The complexity of these trace relations makes it challenging to analytically understand and systematically classify these states. To gain deeper insights into this phenomenon, it is natural to explore other well-studied holographic models. This paper focuses on the D1-D5 CFT, which is central to ${\rm AdS}_3\times {S}^3$ holography and offers a more accessible framework for investigating finite $N$ effects and the associated fortuitous states. This is also the cradle of the fuzzball program (see \cite{Bena:2022rna} for a review), to which the fortuity conjecture of \cite{Chang:2024zqi} poses a major challenge.\footnote{See \cite{Chen:2024oqv} for more challenges to the fuzzball program from the aspect of chaos and the relation between chaos and fortuity.} We believe that the framework laid out in this paper will shed light on the microscopic nature of the numerous horizonless geometries constructed in \cite{Lunin:2001jy, Lunin:2002iz, Taylor:2005db, Kanitscheider:2007wq, Bena:2015bea, Bena:2016agb, Bena:2017xbt, Bakhshaei:2018vux, Ceplak:2018pws, Heidmann:2019zws, Heidmann:2019xrd, Shigemori:2020yuo}.

String theory on \( {\rm AdS}_3 \times {S}^3 \times T^4 \) is dual to the symmetric orbifold of the \( T^4 \) sigma model with exactly marginal deformations \cite{Maldacena:1997re,Maldacena:1998bw,Maldacena:1999bp,Larsen:1999uk, Eberhardt:2018ouy,Eberhardt:2019ywk}. At the orbifold point, the degeneracy of low-lying states grows exponentially with energy, in contrast to the subexponential growth expected in supergravity \cite{deBoer:1998kjm,Benjamin:2015vkc,Benjamin:2016pil}. However, it is believed that exactly marginal deformations give most of these low-lying states large anomalous dimensions,
rendering the spectrum compatible with the supergravity interpretation\cite{Dijkgraaf:1998gf,Seiberg:1999xz,Gaberdiel:2015uca}.

These anomalous dimensions can be studied using conformal perturbation theory. The leading nontrivial contribution to the lifting matrix arises at second order and can be computed using the method introduced by Gava and Narain \cite{Gava:2002xb, Guo:2020gxm, Guo:2019ady}. This approach leverages the supersymmetry algebra, expressing the lifting matrix \(\Delta\) as an anti-commutator
\ie  
\{Q, Q^\dagger\} = \Delta\,.  
\fe  
The computation then reduces to evaluating the matrix elements of the supercharge \(Q\) at \emph{first} order in conformal perturbation theory, which corresponds to a three-point function calculation at the orbifold point. Other studies have directly examined the lifting matrix at second order in conformal perturbation theory \cite{Benjamin:2021zkn, Hughes:2023apl, Guo:2022ifr, Hughes:2023fot}, which involves calculating four-point functions at the orbifold point. The machinery for computing correlations in the symmetric orbifold theory was laid out in \cite{Lunin:2000yv, Lunin:2001pw} based on the covering space method of \cite{hamidi1987interactions}. In the large $N$ and long cycle limit, it was recently found that the spectrum of the second order lifting matrix exhibits an integrable structure \cite{Gaberdiel:2023lco,Frolov:2023pjw}.
At finite $N$, most of these analyses focus on states with relatively small quantum numbers, since extending these results to states with higher quantum numbers is challenging due to the increased complexity introduced by the higher fractional modes. Similarly, extending these studies to higher-order perturbation theory is difficult as it necessitates the evaluation of higher-point functions.

A review of the D1-D5 CFT and the lifting of BPS states can be found in Section~\ref{sec:review_D1D5}.

For BPS states (states with $\Delta = 0$), the supercharge \( Q \) cohomology offers a simpler framework for studying their lifting, since, by a standard argument from Hodge theory, they are in one-to-one correspondence with $Q$-cohomology classes.
In ${\cal N}=4$ SYM \cite{Kinney:2005ej,Grant:2008sk,Chang:2013fba}, it was argued based on the rigidity of $Q$-cohomology that the BPS states at one loop remain BPS to all loops, perhaps even non-perturabtively.
While we do not have a rigorous argument, the relative rigidity of $Q$-cohomology encourages us
to propose the following conjecture:

\begin{conj}[Non-renormalization]\label{conj:BPS_spectrum}
The spectrum of BPS states in the D1-D5 CFT is determined exactly by the lifting matrix at second order, without any higher-order corrections.
\end{conj}

\noindent A similar phenomenon was
observed in the K3 CFT \cite{Keller:2019suk}. 

To gain more evidence for this conjecture, in Section~\ref{sec:lifting_as_cohomology}, we systematically study the (finite $N$) supercharge $Q$-cohomology under first-order conformal perturbation theory. We lay out the general machinery in Sections~\ref{sec:Q-action}. In Section~\ref{sec:N=2cohomology}, we focus on the $N=2$ theory, and find explicit cohomology classes for small charges, whose degeneracy precisely match the known BPS partition function, which was computed using a bootstrap method that assumed no symmetry enhancement under generic deformations \cite{Ooguri:1989fd,Benjamin:2021zkn}. The study of cohomology classes in the $N=3$ theory will be presented in a follow-up work \cite{Chang:2025wgo}.

A significant simplification of the supercharge \( Q \)-cohomology in the D1-D5 CFT compared to \({\cal N}=4\) SYM is that the trace relations in \({\cal N}=4\) SYM are replaced by much simpler relations: the stringy exclusion principle \cite{Maldacena:1998bw}. In the symmetric orbifold theory, the twisted sectors are labeled by cycle shapes, and the stringy exclusion principle eliminates cycle shapes where the total length of nontrivial cycles exceeds \( N \). 
Following \cite{Chang:2024zqi}, the supercharge \( Q \)-cohomology classes are classified into two categories: the monotone classes and fortuitous classes, where the quotient map of the holographic covering is given by the stringy exclusion principle \cite{Maldacena:1998bw}. However, due to the non-commutativity between the stringy exclusion principle projection and the supercharge action, we need to adopt a generalized definition for them introduced in \cite{Chang:2024lxt}.\footnote{In the first version of this paper, this non-commutativity was not recognized, leading to the misidentification of several monotone classes as fortuitous, including the singleton states recently studied in \cite{Hughes:2025car}. We thank Marcel R. R. Hughes and Masaki Shigemori for detailed discussions on this point.} A new feature of this generalized definition is that monotone classes might not always arise by imposing the stringy exclusion principle on the large \( N \) cohomology classes. We refer to those that do have a large $N$ origin as {\it absolute} monotone (see Section~\ref{sec:defns}). The space of fortuitous classes is as defined originally in \cite{Chang:2024zqi} as the quotient of the $Q$-cohomology by the subspace of monotone classes. We apply the generalized definition to classify the cohomology classes in the $N=1$ theory, and show that all the classes are absolute monotone (see Section~\ref{sec:N=1classification}). We also discussed how nontrivial monotone cohomology classes at larger 
$N$ can be projected out at smaller $N$ by the stringy exclusion principle. The study of fortuity in the $N=2$ theory will be presented in the follow-up work \cite{Chang:2025wgo}.

A series of conjectures regarding the bulk duals of the monotone and fortuitous cohomology classes was proposed in \cite{Chang:2024zqi}. In the D1-D5 context, the key statements include the following:
\begin{itemize}
\item The monotone cohomology classes correspond to bulk states arising from the quantization of smooth horizonless geometries \cite{Rychkov:2005ji,Krishnan:2015vha,Shigemori:2019orj,Mayerson:2020acj}, specifically the Lunin-Mathur geometries \cite{Lunin:2001jy, Lunin:2002iz, Taylor:2005db, Kanitscheider:2007wq} and the superstrata geometries \cite{Bena:2015bea, Bena:2016agb, Bena:2017xbt, Bakhshaei:2018vux, Ceplak:2018pws, Heidmann:2019zws, Heidmann:2019xrd, Shigemori:2020yuo}. 
\item The fortuitous cohomology classes correspond to typical microstates of the 3-charge (D1-D5-P) black holes \cite{Strominger:1996sh, Breckenridge:1996is}.

\end{itemize}
\noindent We leave the investigation of these conjectures and the study of large-$N$ cohomology for the future work.

In Section~\ref{sec:prod_coho}, we construct two-cycle composite BPS states by taking products of two single-cycle BPS states (of cycle length $w_1$ and $w_2$, respectively) and performing certain projections. We further extend the discussion to multi-cycle composite BPS states in  section~\ref{sec_multi_cycle}. Depending on whether each constituent is monotone or fortuitous, such states have different gravitational interpretations, and are the subject of Section~\ref{sec_two_center_massive}. 

For these two-cycle composites, when both constituents are fortuitous, they naturally have the interpretation of black hole bound states or near horizon geometries of two-centered black hole solutions \cite{Giusto:2004id,Giusto:2004ip,deBoer:2008fk}. Instead, if one is fortuitous and the other monotone, then in the limit \( w_2 \gg w_1 \), the composite can be interpreted as a fortuitous excitation probing a monotone background. A length-\( w_1 \) fortuitous state in the large $N$ limit is dual to a massive stringy excitation in the vacuum \({\rm AdS}_3 \times {S}^3 \times T^4\). Thus, it is natural to propose that the composite state, in the large $N$ limit (\( N = w_1 + w_2 \gg w_1  \)), is dual to a BPS stringy excitation on a smooth horizonless geometry corresponding to the cycle-length \( w_2 \) monotone state. It is surprising that the non-BPS stringy excitations in ${\rm AdS}_3\times {S}^3\times T^4$ vacuum can become BPS in smooth horizonless geometries, such as the Lunin-Mathur and superstrata geometries. It would be interesting to study the worldsheet string theory on such geometries following \cite{Martinec:2017ztd
,Martinec:2018nco}, and check if there exist stringy modes that satisfy the BPS condition.

\section{Review of the D1-D5 CFT}\label{sec:review_D1D5}

The D1-D5 CFT describes the low energy limit of a supersymmetric configuration of D1- and D5-branes in type IIB string theory on $\bR^6 \times \cM$, where $\cM$ can be $T^4$ or K3. Its conformal moduli correspond to the geometric and flux deformations.

\subsection{Symmetric orbifold of the $T^4$ CFT}\label{sec:sym_orb}

The $T^4$ CFT is a system of four free bosons, four free left-moving fermions, and four free right-moving fermions,
\ie\label{eqn:T4_fields}
X^\mu(z,\bar z)\,,\quad \psi^{\pm}(z)\,,\quad \bar\psi^{\pm}(z)\,,\quad \tilde\psi^{\pm}(\bar z)\,,\quad \tilde{\bar\psi}^{\pm}(\bar z)\,,
\fe
where $\mu=1,\,\cdots,\,4$.
We will mainly consider the theory in the Neveu–Schwarz (NS) sector, where the fermions carry half-integer modes. The free bosons $X^\mu$ are valued in $T^4$, i.e. satisfy the identification $X^\mu\sim X^\mu+2\pi R^\mu$, where $R^\mu$ are the radii of $T^4$. Let us denote the Hilbert space in the NS sector by ${\cal H}$, which is simply the Fock space generated by the free boson and fermion creation operators acting on the NS sector ground state $\ket 1$.

Now, consider the symmetric orbifold theory ${\rm Sym}^N(T^4)$, which contains an untwisted sector and several twisted sectors. The Hilbert space of the untwisted sector is a (graded) symmetric tensor product of ${\cal H}$,
\ie
S^N\cH=\Big(\underbrace{\cH\otimes\cH\otimes\cdots\otimes\cH}_{N\,{\rm times}}\Big)^{S_N}\,.
\fe 
The twisted sectors correspond in one-to-one to the conjugacy classes of $S_N$. Let us introduce the notation for a cycle shape $p$ (an integer partition of $N$),
\ie\label{eqn:p_partition}
p=(\underbrace{1,1,\cdots,1}_{N_1\,{\rm times}},\underbrace{2,2,\cdots,2}_{N_2\,{\rm times}},\cdots)=(1^{N_1},2^{N_2},\cdots)\,,\quad N=N_1+2N_2+3N_3+\cdots\,,
\fe
which specifies a conjugacy class consisting of elements of $S_N$ with $N_1$ one-cycles, $N_2$ two-cycles, and so on.

The untwisted sector is the twisted sector defined above with $N_1=N$. The Hilbert space $\cH_p$ of a twisted sector $p$ can be explicitly constructed as follows. Given a group element $g\in S_N$ belonging to the conjugacy class specified by the cycle shape $p$, let ${\cal H}_g$ denote the Hilbert space obtained by quantizing the tensor product theory $(T^4)^{N}$ on ${S}^1$ with a twisted boundary condition
\ie
X^{[i]}(\sigma+2\pi) = X^{[g(i)]}(\sigma)\,,\quad \psi^{[i]}(\sigma+2\pi) = \psi^{[g(i)]}(\sigma)\,,
\fe
where $X^{[i]}$ and $\psi^{[i]}$ denote the free fields in \eqref{eqn:T4_fields} in the $i$-th $T^4$ CFT, and $\sigma\in[0,2\pi)$ is the coordinate of the ${S}^1$. The Hilbert space $\cH_p$ is defined as the direct sum of $\cH_g$ over all group elements belonging to the conjugacy class (cycle shape) $p$,
\ie\label{eqn:Hp}
\cH_p=\bigoplus_{g\in p}\cH_g\,.
\fe

The Hilbert space ${\cal H}_g$ factorizes into a tensor product of single-cycle Hilbert spaces. For example, consider the group element\footnote{One should be careful not to confuse the notation \eqref{eqn:ge_ex} for group elements and that \eqref{eqn:p_partition} for conjugacy classes (cycle shapes).}
\ie\label{eqn:ge_ex}
g=(1,\cdots,w_1)(w_1+1,\cdots,w_1+w_2)\cdots(N-w_\nu+1,\cdots,N)\in S_N\,.
\fe
The Hilbert space $\cH_g$ factorizes as 
\ie
\cH_g=\bigotimes_{n=1}^\nu \cH_{(w_n)}\,,
\fe
where $\cH_{(w)}$ is the Hilbert space of the single-copy $T^4$ CFT on a circle with $w$ times the radius (in particular, $\cH_{(1)}=\cH$). More explicitly, the Hilbert space $\cH_{(w)}$ is a Fock space generated by the bosonic and fermionic creation operators with fractional modes,
\ie\label{eqn:frac_modes}
&\alpha^{(w)}_{n+\frac{m}{w}} 
= \frac{1}{\sqrt w}\oint_{z=0} \sum^w_{j=1}e^{-2\pi i \frac{m}{w}j}X^{[j]}(z)\, z^{n+\frac{m}{w}} \, dz , 
\\
&\psi^{(w)}_{r+\frac{m}{w}} = \frac{1}{\sqrt w}\oint_{z=0} \sum^w_{j=1}e^{-2\pi i \frac{m}{w}j}\psi^{[j]}(z) \, z^{r+\frac{m}{w}-\frac{1}{2}} \, dz\, .
\fe 
The bosons $X^{[j]}$ satisfy the boundary conditions $X^{[j]}(e^{2\pi i}z)\sigma_w(0)=X^{[j+1]}(z)\sigma_w(0)$, where $j=1,\,\cdots,\,w$ with the identification of $w+1 \sim 1$, and similarly for the fermions $\psi^{[j]}$. Under the state/operator correspondence, the ground state in $\cH_{(w)}$ is dual to a ${\mathbb Z}_w$ twisted operator $\sigma_w$, and $\cH_{(w)}$ can be equivalently regarded as the radially quantized Hilbert space with $\sigma_w$ inserted at the origin.

Given $\ket{\Psi}\in \cH_g$, we obtain a state $\ket{\{\Psi\}}\in\cH^{S_N}_p$ by summing over $S_N$:
\ie\label{eqn:SN_symmetrization}
\ket{\{\Psi\}} = \sum_{g\in S_N}g\ket{\Psi}\in \cH^{S_N}_p\,.
\fe

For studying the supercharge cohomology, it is convenient to choose the normalization
\ie\label{eqn:normalization}
\bra\Psi\ket{\Psi}=1\,,
\fe
and hence the state $\ket{\{\Psi\}}$ has the normalization
\ie
\bra{\{\Psi\}}\ket{\{\Psi\}}=N!\,.
\fe

The full Hilbert space of the free, undeformed D1-D5 CFT of rank $N$ is a direct sum over the Hilbert spaces of all twisted sectors,
\ie\label{eqn:full_Hilbert_space}
\cH_N=\bigoplus_{p} \cH^{S_N}_p \,,
\fe
where $\cH^{S_N}_p$ is the $S_N$ invariant subspace of $\cH_p$. There is an isomorphism
\ie\label{hilbert_space_isomorphism}
    \iota: \quad \cH^{S_N}_{p = (1^{N_1},\,2^{N_2},\,\cdots)} \cong \bigotimes_{w=1}^\infty S^{N_w}\cH^{{\mathbb Z}_w}_{(w)}\,,
\fe
where $\cH^{{\mathbb Z}_w}_{(w)}$ is the ${\mathbb Z}_w$-invariant sector in $\cH_{(w)}$, i.e. the sector with integer (resp. half-integer) spins $\ell = h-\tilde h $ for the bosonic (resp. fermionic) states.

The inverse map $\iota^{-1}$ of the Hilbert space isomorphism \eqref{hilbert_space_isomorphism} is given by summing over the $S_N$ orbit. By definition, orbifold/gauging restricts to the $S_N$ invariant subspace, which is isomorphic to the $S_N$ orbit,\footnote{This isomorphism is a general fact about $G$-spaces when $G$ is finite.} and hence the inverse map is surjective. It is also injective by noting that the $S_N$ orbit of an arbitrary state on the right consists of mutually orthogonal elements.

Given the abundance of Hilbert spaces introduced in this section, a summary is provided in Table~\ref{tab:H} for the readers' convenience.
\begin{table}[H]
    \centering
    \begin{tabular}{|c|c|}
        \hline 
        $\cH$ & single-copy CFT Hilbert space
        \\
        $\cH_{(w)}$ & single-copy CFT on circle with $w$ times the radius
        \\
        $\cH_g = \bigotimes_{n=1}^\nu \cH_{(w_n)}$ & Hilbert space from $g$-twisted quantization
        \\
        $\cH_p = \bigoplus_{g \in p} \cH_g$ & formal direct sum over elements of conjugacy class $p$
        \\
        $\cH_p^{S_N} = \cH_p/S_N$ & twisted sector associated with conjugacy class $p$
        \\
        $\cH_N = \bigoplus_p \cH_p^{S_N}$ & Hilbert space of free D1-D5 CFT (symmetric product orbifold)
        \\\hline
    \end{tabular}
    \caption{Glossary of Hilbert spaces.}
    \label{tab:H}
\end{table}

\subsection{BPS states at the orbifold point} 
\label{sec:free-BPS}
The $T^4$ CFT has the contracted large ${\cal N}=4$ superconformal symmetry \cite{ali1993new} with central charge $c=6$, generated by the left-moving currents 
\ie\label{eqn:TGK}
T\,,\quad G^{\pm}\,,\quad G'^{\pm}\,,\quad J^\pm\,,\quad J^3\,,
\fe
defined in appendix  \ref{app: coveringspace},
and 
\ie\label{eqn:alpha_psi}
\alpha^i=\partial(X^{2i-1}+iX^{2i})\,,\quad \bar\alpha^i=\partial(X^{2i-1}-iX^{2i})\,,\quad \psi^\pm\,,\quad \bar\psi^\pm\,,
\fe
where $i=1,2$. Here, $T(z)$ is the stress tensor, $G^{\pm}(z), G'^{\pm}(z)$ are the four superconformal currents, $J^\pm(z), J^3(z)$ are the spin-1 currents that generate the $\mathfrak{su}(2)$ Kac-Moody algebra at level 1, $\alpha^i(z), \bar\alpha^i(z)$ are the spin-1 currents that generate the $\mathfrak{u}(1)^4$ Kac-Moody algebra, and finally $\psi^\pm(z), \bar\psi^\pm (z)$ are the spin-$\frac12$ free fermions.  The pairs $(G^{'+}, G^-)$ and $(G^+,G^{'-})$ form doublets under SU$(2)_R$ respectively. Setting $\alpha^i$, $\psi^\pm$, and $\bar\psi^\pm$ to zero is a consistent truncation that recovers the small $\cN=4$ algebra. The $T^4$ CFT possesses an additional SU(2) symmetry that acts as an outer automorphism of the small $\cN=4$ algebra. We denote its generators by 
$K^{\pm}, K^3$. In our convention, the fermions $\psi^+$ and $\psi^-$ carry automorphism charge $-\tfrac{1}{2}$, while $\bar\psi^+$ and $\bar\psi^-$ carry automorphism charge $+\tfrac{1}{2}$. 
 The additional $\mathfrak{su}(2)$ currents are composites of the free bosons and fermions in \eqref{eqn:alpha_psi} with the explicit formulae given in \eqref{eqn:TGK_in_alpha-psi} and \eqref{eqn:outer_su2}. Throughout, we label right-moving currents with a tilde.

The symmetric orbifold theory inherits the superconformal symmetry with the generators given by the sum of the generators over all copies; for example,
\ie
T=\sum_{i=1}^N T^{[i] }\,,
\fe
and similarly for the other currents.

The \emph{superconformal primary} states (in the NS sector) are defined as the states annihilated by all the positive modes of the currents \eqref{eqn:TGK}, \eqref{eqn:alpha_psi} and their right-moving counterparts. Let us focus on the right-moving part. The (right-moving) \emph{chiral primary} states are the superconformal primaries further annihilated by the supercharges
\ie\label{eqn:chiral_primary_cond}
Q:=\widetilde G^{+}_{-\frac12}\,,\quad Q':=\widetilde G'^{+}_{-\frac12}\,. 
\fe
Together with their Hermitian conjugates $Q^\dagger = \widetilde G^-_{\frac12}$ and $Q'^\dagger = \widetilde G'^-_{\frac12}$, they satisfy the supersymmetry algebra
\ie\label{eqn:SUSY_alg}
\{Q,Q^\dagger\}=\{Q',Q'^\dagger\}=2(\widetilde L_0-\widetilde K_0^3) =: \Delta\,,
\fe
and the BPS condition
\ie\label{eqn:BPS_cond}
\tilde h=\tilde j\,,
\fe
where $\tilde j$ is the eigenvalue of $\widetilde K^3_0$.

Since the $\mathfrak{u}(1)^4$ charges do not appear in the BPS formula \eqref{eqn:BPS_cond}, for a generic $T^4$, the states with nonzero momentum or winding necessarily violate the BPS condition; in other words, momentum and winding increase $\tilde h$ but leave $\tilde j$ intact.\footnote{For special $T^4$ moduli, there can be states with nonzero momentum and winding, in such a way that $h$ is increased but not $\tilde h$, so that the states remain BPS on the right.}
As we are interested in BPS states, we can and will henceforth restrict ourselves to the sector with zero momentum and winding.

In the contracted large ${\cal N}=4$ superconformal algebra, besides the supercharges $Q$ and $Q'$, there are four spin-$\frac12$ fermionic generators that satisfy the BPS condition \eqref{eqn:BPS_cond} and generate a Clifford algebra
\ie\label{eqn:Clifford}
\{\tilde\psi^+_{-\frac12},\tilde{\bar\psi}^-_{\frac 12} \}=-1\,,\quad \{\tilde{\bar\psi}^+_{-\frac 12},\tilde{\psi}^-_{\frac12}\}=1\,.
\fe
In the zero momentum and winding sector, the supercharges $Q$, $Q'$, $Q^\dagger$, and $Q'^\dagger$ commute with this Clifford algebra,
\ie\label{eqn:Q_Clifford_commute}
&\{Q,\tilde \psi_{-\frac12}^+ \}= \{Q,\tilde {\bar \psi}_{-\frac12}^+ \}=\{Q',\tilde \psi_{-\frac12}^+ \}= \{Q',\tilde {\bar \psi}_{-\frac12}^+ \}=0\,,
\\
&\{Q,\tilde \psi_{\frac12}^- \}=-\tilde\alpha^2_0\,,\quad \{Q,\tilde {\bar \psi}_{\frac12}^- \}=\tilde{\bar\alpha}^2_0\,,
\quad \{Q',\tilde \psi_{\frac12}^- \}=\tilde\alpha^1_0\,,\quad \{Q',\tilde {\bar \psi}_{\frac12}^- \}=-\tilde{\bar\alpha}^1_0\,,
\fe
and analogously for $Q^\dag$ and $Q'^\dag$ by taking the Hermitian conjugate of \eqref{eqn:Q_Clifford_commute}.\footnote{The bosonic generators $K^{+}_{-1}$, $K^{-}_{1}$ and $K^3_0-\frac c{12}$ form a ${\rm SU}(2)$ Lie algebra and satisfy the BPS condition. However, $K^{-}_{1}$ does not commute with the supercharge $Q$; hence, we cannot use this ${\rm SU}(2)$ to organize the $Q$-cohomology.
}
Therefore, the \(\tilde\psi^+_{-\frac{1}{2}}\) and \(\tilde{\bar{\psi}}^+_{-\frac{1}{2}}\) descendants of a chiral primary state also satisfy the BPS condition, forming a four-dimensional representation (quartet) of the Clifford algebra \eqref{eqn:Clifford}.

The \emph{chiral-chiral primary} states are the superconformal primaries further annihilated by both \eqref{eqn:chiral_primary_cond} and their left-moving counterparts. The only chiral-chiral primary state in the $T^4$ CFT is the vacuum state with dimension $h=\tilde h=0$. The chiral-chiral primary state in the single-cycle twisted sector $\cH^{{\mathbb Z}_w}_{(w)}$ is the vacuum state of the single-copy theory on a circle with $w$ times the radius, giving weights and (R-symmetry) spins
\ie
h=\tilde h=\frac{w-1}{2}=j=\tilde j\,.
\fe
It also carries vanishing automorphism charges, $K=\tilde K=0$, in both the left- and right-moving sectors. The chiral-chiral primary states in the multi-cycle twisted sectors are exhaustively given by the products of the chiral-chiral primary states in the single-cycle twisted sectors; in particular, they are all Lorentz scalars. 

In the single-cycle twisted sector $\cH^{{\mathbb Z}_w}_{(w)}$, the chiral-chiral primary states and their $\psi^{+(w)}_{-\frac12}$, $\bar\psi^{+(w)}_{-\frac 12}$, $\tilde\psi^{+(w)}_{-\frac12}$, $\tilde{\bar\psi}^{+(w)}_{-\frac 12}$ descendants give the full set of $\frac12$-BPS states satisfying the BPS condition \eqref{eqn:BPS_cond} and the left-moving BPS condition\footnote{The superscript $(w)$ was introduced in \eqref{eqn:frac_modes}.}
\ie\label{eqn:hol_BPS_cond}
h=j\,.
\fe
They can be organized by two Clifford algebras: \eqref{eqn:Clifford} and its left-moving counterpart, and be denoted by
\ie\label{eqn:w-BPS}
\ket{w_{\pm\pm,\pm\pm}}\,,
\fe
where the first $\pm\pm$ denote the quartet of the left-moving Clifford algebra, and the second $\pm\pm$ denote the quartet of the right-moving Clifford algebra. More precisely, the common bottom component $\ket{w_{--,--}}$ of the two quartets is the chiral-chiral primary state, 
and the other components are
\ie
\ket{w_{a_1 a_2,\bar a_1\bar a_2}} = (\psi^{+}_{-\frac{1}{2}})^{\frac{1a_11}2}(\bar{\psi}^{+}_{-\frac{1}{2}})^{\frac{1a_21}2}(\tilde{\psi}^{+}_{-\frac{1}{2}})^{\frac{1\bar a_11}2}(\tilde{\bar{\psi}}^{+}_{-\frac{1}{2}})^{\frac{1\bar a_21}2}\ket{w_{--,--}}\,,
\fe
where $1a1$ means $1\pm1$ for $a=\pm$.
A {\it trivial cycle} is defined to be the ground state $\ket{1_{--,--}}$ in the single-cycle Hilbert space.  A general multi-cycle $\frac12$-BPS state is given by products of single-cycle $\frac12$-BPS states. 

The $\frac14$-BPS single-cycle states satisfying the right-moving BPS condition \eqref{eqn:BPS_cond} can be constructed by acting the left-moving operators on the $\frac12$-BPS states. Let us choose a basis for this $\frac14$-BPS single-cycle space as
\ie
\ket{w_{i,\alpha}}\,,
\fe
where $\alpha=\pm\pm$ and the index $i$ labels the different combinations of left-moving generators acting on $\ket{w_{--,\pm\pm}}$. The explicit R-charges of the states $\ket{w_{i,\alpha\beta}}$
are
\ie\label{bps_clifford_quartet}
\text{BPS Clifford quartet}: \quad
    \begin{bmatrix}
        \ket{w_{i,++}}
        \\
        \ket{w_{i,+-}}
        \\
        \ket{w_{i,-+}}
        \\
        \ket{w_{i,--}}
    \end{bmatrix},
    \quad 
    \tilde j =
    \begin{bmatrix}
        \frac{w+1}{2}
        \\
        \frac{w}{2}
        \\
        \frac{w}{2}
        \\
        \frac{w-1}{2}
    \end{bmatrix}.
\fe

Let us consider a multi-cycle $\frac14$-BPS state, which can be obtained by starting with a tensor product of single-cycle $\frac14$-BPS states:
\ie\label{eqn:1/4-BPS_mc_pre}
&\big|1^{N_{1,1}}_{i_{1,1},\alpha_{1,1}},1^{N_{1,2}}_{i_{1,2},\alpha_{1,2}} ,\cdots, 2^{N_{2,1}}_{i_{2,1},\alpha_{2,1}},\cdots,3^{N_{3,1}}_{i_{3,1},\alpha_{3,1}},\cdots \big\rangle
\\
&:=\left(\big|1^{N_{1,1}}_{i_{1,1},\alpha_{1,1}} \big\rangle \otimes \big| 1^{N_{1,2}}_{i_{1,2},\alpha_{1,2}} \big \rangle \otimes \cdots\right)\otimes \left(\big| 2^{N_{2,1}}_{i_{2,1},\alpha_{2,1}}  \big \rangle\otimes\cdots\right) \otimes\left(\big|3^{N_{3,1}}_{i_{3,1},\alpha_{3,1}} \big \rangle\otimes\cdots\right)\otimes\cdots
\\
&
\in \bigotimes_{n=1}^\nu \cH_{(w_n)}^{\otimes N_n}\,,
\fe
where $N_{m,n}$ is the number of cycles with the same length $m$ and excitations labeled by $i_{m,n},\alpha_{m,n}$, and $\sum_{n}N_{m,n}=N_m$, $\sum_{m=1}^\infty m N_m 
= N$. After summing over $S_N$
as in \eqref{eqn:SN_symmetrization}, we obtain a $\frac14$-BPS state in the twisted sector:
\ie\label{eqn:1/4-BPS_mc}
\big|\{1^{N_{1,1}}_{i_{1,1},\alpha_{1,1}},1^{N_{1,2}}_{i_{1,2},\alpha_{1,2}} ,\cdots, 2^{N_{2,1}}_{i_{2,1},\alpha_{2,1}},\cdots \}\big\rangle\in \cH^{S_N}_p\,,\quad p=(1^{N_1},2^{N_2},\cdots)\,.
\fe

\ie
V^{S_N}_{p} & :=\underset{i_{m,n},\alpha_{m,n}}{\rm span}\left(
\big|\{1^{N_{1,1}}_{i_{1,1},\alpha_{1,1}},1^{N_{1,2}}_{i_{1,2},\alpha_{1,2}} ,\cdots, 2^{N_{2,1}}_{i_{2,1},\alpha_{2,1}},\cdots \}\big\rangle
\right)\in \cH^{S_N}_p\,,
\fe
where $p=(1^{N_1},2^{N_2},\cdots)$. We will refer to the space $V_N$ after direct summing over $p$ as the {\it free} BPS sector---the BPS sector in the free symmetric orbifold theory,\footnote{The free BPS sector here is analogous to the classically-BPS sector in the $\cN=4$ SYM \cite{Chang:2022mjp}.}
\ie
V_N& =\bigoplus_{\text{cycle shapes }p} V_p^{S_N}\,.
\fe

The space $V_{p}^{S_N}$ forms a representation of $(\mathrm{Cliff})^N$. However, the exactly marginal deformation discussed in Section~\ref{sec_review_lifting} only preserves the diagonal Clifford algebra generated by the summations
\ie\label{eqn:diag_clif}
\psi^{\rm diag}_r=\frac{1}{\sqrt{N}}\sum_{j=1}^{N} \psi^{[j]}_r\,.
\fe
For simplicity, in the case without identical cycles (i.e. $N_n=1$), we have
\ie
\psi^{\rm diag}_r=\sum_{n=1}^\nu \sqrt\frac{w_n}{N}\psi^{(w_n)}_r\,.
\fe
Here, $\psi$ denotes all the fermions $\tilde\psi^\pm$ and $\tilde{\bar \psi}^\pm$.
Since the diagonal Clifford algebra \eqref{eqn:Clifford} commutes with $Q$ and $Q^\dagger$, we focus on the top subspace given by
\ie\label{eqn:top_space}
V^{\rm top}_N=\tilde\psi^{+\rm diag}_{-\frac12}\tilde{\bar\psi}^{-\rm diag}_{\frac12}\tilde{\bar\psi}^{+\rm diag}_{-\frac12}\tilde\psi^{-\rm diag}_{\frac12}V_N\,.
\fe
Note that here we project onto the top components of the diagonal Clifford algebra, but other non-top components in individual cycles can still show up. For later use, we also define
\ie
V_g:=\cH_g\big|_{\tilde h=\tilde j}\,.
\fe
For example, the state \eqref{eqn:1/4-BPS_mc_pre} is inside $V_g$:
\ie
&\big|1^{N_{1,1}}_{i_{1,1},\alpha_{1,1}},1^{N_{1,2}}_{i_{1,2},\alpha_{1,2}} ,\cdots, 2^{N_{2,1}}_{i_{2,1},\alpha_{2,1}},\cdots,3^{N_{3,1}}_{i_{3,1},\alpha_{3,1}},\cdots \big\rangle\in V_g
\fe
with the group element $g\in S_N$ explicitly given by
\ie
g&=\left[(1)(2)\cdots(N_1)\right]\left[(N_1+1,N_1+2)\cdots((N_1+2N_2-1, (N_1+2N_2)\right]
\\
&\quad \times \left[(N_1+2N_2+1,N_1+2N_2+2,N_1+2N_2+3)\cdots\right]\cdots\,.
\fe

\subsection{Lifting under conformal perturbation theory}
\label{sec_review_lifting}

Consider deforming the symmetric product theory by an exactly marginal operator $\Phi(z,\bar z)$,
\be\label{eqn:deformation}
S = S_{Sym^N(T^4)} + g \int  d^2z \, \Phi(z,\bar z).
\ee
We choose $\Phi(z,\bar z)$ to be the state/operator dual of
\be\label{eqn:exactly_marginal_deformation}
\frac{i}{\sqrt{2}}(G^-_{-\frac{1}{2}}\widetilde{G}^{'-}_{-\frac{1}{2}}-G^{'-}_{-\frac{1}{2}}\widetilde{G}^{-}_{-\frac{1}{2}})\ket{\{1_{--,--}^{N-2},2_{--,--}\}},
\ee
where $\ket{\{1_{--,--}^{N-2},2_{--,--}\}}$ contains $N-2$ trivial cycles and the bottom component of the $\frac12$-BPS quartet in the ${\mathbb Z}_2$ twisted sector, with $h=j=\tilde{h}=\tilde{j}=\frac{1}{2}$. 
In the AdS/CFT correspondence, this particular choice of the exactly marginal operator corresponds to tuning the RR flux in the $\mathrm{AdS}_3 \times \mathrm{S}^3$ factor \cite{Gaberdiel:2023lco}.

The \( \frac{1}{2} \)-BPS states are protected under any (contracted large) $\cN = (4,4)$-preserving exactly marginal deformation. To see this, note that their lifting would require the recombination of a pair of chiral-chiral primary states whose spins differ by a half. However, in the $T^4$ symmetric orbifold theory, all chiral-chiral primary states in the theory are scalars, hence such a recombination is not possible. Some \( \frac{1}{4} \)-BPS states are also protected, while the others acquire anomalous dimensions. A central question in understanding D1-D5 black hole microstates is to describe these unlifted states.

The anomalous dimension is a sum of the left-moving and right-moving anomalous dimensions, \( \delta h + \delta \tilde h \). The deformation given in \eqref{eqn:deformation} and \eqref{eqn:exactly_marginal_deformation} preserves Lorentz symmetry; hence, the  ``anomalous spin" \( \delta \ell = \delta \tilde{h} - \delta h \) must be zero. Therefore, it suffices to compute the right-moving anomalous dimension $\delta\tilde h$. In this paper, we will focus on the lifting of states in the free BPS sector $V_N$. For a state $\ket\Psi\in V_N$ satisfying \eqref{eqn:BPS_cond} in the symmetric orbifold theory, we have
\be\label{anomalous_dimension}
\delta \tilde{h} \ket{\Psi} = (\widetilde{L}_0 - \widetilde{K}_0^3) \ket{\Psi} = \frac12 \{ Q^\dagger, Q \} \ket{\Psi} = \frac12 \{ Q'^\dagger, Q' \} \ket{\Psi}\,,
\ee
where $Q$, $Q'$, and their Hermitian conjugates have vanishing zeroth-order actions on $\ket\Psi$.

 In conformal perturbation theory, the generators of the contracted large ${\cal N}=4$ superconformal algebra admit a perturbative expansion in the deformation parameter $g$, e.g.
\ie\label{L_-2_expansion}
(L_{-2})_0+g (L_{-2})_1 + g^2 (L_{-2})_2 +\cdots\,,
\fe
where $(L_{-2})_n$ is defined by its matrix elements 
\ie\label{eqn:n-order-deformed_L-2}
\langle \Psi_i | (L_{-2})_n |\Psi_j\rangle := \langle \Psi_i | L_{-2} \left(\int d^2z \, \Phi(z,\bar z)\right)^n |\Psi_j\rangle\,.
\fe
Simple dimensional analysis shows that at any fixed order in conformal perturbation theory, each term like $(L_{-2})_n$ has a definite eigenvalue of the undeformed scaling dimension $\Delta_0:=(L_0+\widetilde L_0)$, e.g.
\ie
[\Delta_0,(L_{-2})_n]=2(L_{-2})_n\,.
\fe
Consequently, the right-moving supercharges $Q_n$, $Q'_n$, $Q^\dagger_n$, $Q'^\dagger_n$ as well as the finite-order deformed left-moving contracted large ${\cal N}=4$ superconformal algebra generators commute with the free BPS bound, and hence preserve the free BPS sector defined in Section~\ref{sec:free-BPS}.

\subsection{Modified index}

The fact that the contracted large $\cN=4$ superconformal algebra contains 4 pairs of free bosons and fermions suggests that any theory enjoying this symmetry contains a copy of the free $T^4$ sigma model. The presence of free fermions then renders the standard supersymmetric indices vanishing. To obtain meaningful indices, one way is to decouple the free $T^4$ sigma model and study the quotient theory, which has a different symmetry algebra (see \cite{Gukov:2004ym,Gukov:2004fh} for discussions in a related model). Instead, we follow the classic treatment \cite{Maldacena:1999bp} to define and compute a modified index.\footnote{Performing the $T^4$ quotient is akin to studying the 4d $\cN=4$ super-Yang-Mills with gauge group $SU(N)$ instead of $U(N)$, by factoring out the ``center-of-mass'' $U(1)$ free theory.}

The BPS states satisfying \eqref{eqn:BPS_cond} are in the one-dimensional short multiplet of the supersymmetry algebra \eqref{eqn:SUSY_alg}, while the non-BPS states form four-dimensional long multiplets. Hence, we can construct an index that counts the number of short multiplets. Let us start with the NS sector partition function (with $(-1)^F$ inserted in the trace),
\ie
Z_{\rm NS}(\tau,\bar\tau,z,\bar z)=\Tr_{\rm NS}\left[(-1)^F q^{L_0-\frac c{24}}\bar q^{\widetilde L_0-\frac c{24}}y^{2K^3_0} \bar y^{2\widetilde K^3_0}\right]\,,
\fe
where $y=e^{2\pi i z}$. The index is defined by
\ie
I'_{\rm NS}(\tau,z)=\bar q^{\frac c{24}}Z_{\rm NS}(\tau,\bar\tau,z,\bar z)\big|_{\bar z=-\frac12\bar \tau}=\Tr_{\rm NS} \left[(-1)^F q^{L_0-\frac c{24}}y^{2K^3_0}\right]\,.
\fe
However, because the BPS states form quartets under the Clifford algebra \eqref{eqn:Clifford} from the free fermions $\psi^\pm$ and $\bar\psi^\pm$, the index $I'_{\rm NS}$ always vanishes due to the cancellation between bosons and fermions.

To get a non-vanishing counting, we consider the modified index
\ie
I_{\rm NS}(\tau,z)&=\frac12 \Tr_{\rm NS}\left[(-1)^F \left(2\widetilde K^3_0\right)\left(2\widetilde K^3_0-1\right)q^{L_0-\frac c{24}}y^{2K^3_0}\right]
\\
&=\frac12 \bar q^{\frac c{24}-1}\partial_{\bar y}^2Z_{\rm NS}(\tau,\bar\tau,z,\bar z)\Big|_{\bar z=-\frac12\bar \tau}\,,
\fe
which is related to the modified index $I_R(\tau,z)$ in
\cite{Maldacena:1999bp} by a spectral flow from the NS sector to the Ramond (R) sector,
\ie
I_R(\tau,z)=q^{\frac c{24}}y^{-\frac c6}I_{\rm NS}\left(\tau,z-\frac \tau2\right)\,.
\fe
The modified index in the R-sector can be computed by taking derivatives on the R-sector partition function (with $(-1)^F$ inserted in the trace) as
\ie
I_R(\tau,z)&=\frac12 \partial_{\bar y}^2Z_{\rm R}(\tau,\bar\tau,z,\bar z)\big|_{\bar z=0}\,,
\\
Z_{\rm R}(\tau,\bar\tau,z,\bar z)&=\Tr_{\rm R}\left[(-1)^F q^{L_0-\frac c{24}}\bar q^{\widetilde L_0-\frac c{24}}y^{2J_0} \bar y^{2\widetilde J_0}\right]\,.
\fe
The spectral flow acts on the R- and NS-sector partition functions as
\ie
Z_{\rm R}(\tau,\bar\tau,z,\bar z)=q^{\frac c{24}}y^{-\frac c6}\bar q^{\frac c{24}}\bar y^{-\frac c6} Z_{\rm NS}\left(\tau,\bar\tau,z-\frac\tau2,\bar z-\frac{\bar\tau}2\right)\,.
\fe

Following \cite{Maldacena:1999bp}, let us compute the modified index in the $T^4$ symmetric orbifold theory. We start with the partition function of the $T^4$ CFT in the R-sector
\ie\label{eqn:T4_pf}
Z_{T^4,{\rm R}}=
\left(\frac{\theta_1(z|\tau)}{\eta(\tau)}\right)^2\frac{1}{\eta(\tau)^4}\overline{\left(\frac{\theta_1(z|\tau)}{\eta(\tau)}\right)^2\frac{1}{\eta(\tau)^4}}\,,
\fe
where we have restricted to the zero momentum and winding sector. Let us rewrite the partition function as
\ie\label{eqn:T4_pf_exp}
Z_{T^4,{\rm R}}=\sum_{h,\tilde h,j,\tilde j}c(h,\tilde h,2j,2\tilde j)q^h \bar q^{\tilde h}y^{2j}\bar y^{2\tilde j}\,,
\fe
where the coefficients $c(h,\tilde h,j,\tilde j)$ can be extracted from the explicit formula \eqref{eqn:T4_pf}. Now, let us consider the index
\ie
I_{T^4,{\rm R}}=\frac12 \partial_{\bar y}^2Z_{T^4,{\rm R}}\big|_{\bar y=1}=-\left(\frac{\theta_1(z|\tau)}{\eta(\tau)}\right)^2\frac{1}{\eta(\tau)^4}=\sum_{h,j}\hat c(h,2j)q^h y^{2j}\,,
\fe
where the coefficients $\hat c(h,2j)$ and $c(h,\tilde h,2j,2\tilde j)$ are related by
\ie
\hat c(h,2j)=\frac12 \sum_{\tilde j}(2\tilde j)^2c(h,0,2j,2\tilde j)\,.
\fe

By the Dijkgraaf-Moore-Verlinde-Verlinde (DMVV) formula \cite{Dijkgraaf:1996xw}, the grand partition function of the $T^4$ symmetric orbifold theories is given by
\ie
{\cal Z}_{\rm R}=\sum^{\infty}_{k=0}p^k Z_{{\rm Sym}^k(T^4),{\rm R}}=\prod^\infty_{n=1}\prod_{\substack{h,\tilde h,j,\tilde j \\ h-\tilde h\in n{\mathbb Z}}}\frac{1}{(1-p^n q^{\frac{h}{n}} \bar q^{\frac{\tilde h}{n}} y^{2j} \tilde y^{2\tilde j})^{c(h,\tilde h,2j,2\tilde j)}}\,,
\fe
where $c(h,\tilde h,2j,2\tilde j)$ is the coefficient in the single-copy partition function \eqref{eqn:T4_pf_exp}. Now, let us compute the grand index
\ie\label{eqn:R_modified_index}
{\cal I}_{\rm R}=\sum_{k=0}^\infty p^k I_{{\rm Sym}^k(T^4),{\rm R}}=\frac12 \partial_{\bar y}^2{\cal Z}_{\rm R}\big|_{\bar y=1}=\sum_{n=1}^\infty \sum_{m=0}^\infty\sum_{j\in \frac12{\mathbb Z}}\frac{\hat c(nm,2j) p^nq^my^{2j}}{(1-p^nq^my^{2j})^2}\,.
\fe
The index in the NS sector can be obtained by the spectral flow from the R sector
\ie\label{eqn:NS_modified_index}
{\cal I}_{\rm NS}(p,q,y)=\sum_{k=0}^\infty p^k I_{{\rm Sym}^k(T^4),{\rm NS}}={\cal I}_{\rm R}(p q^\frac14 y,q,q^{\frac12} y)\,.
\fe
For instance, the expansion of the $N=2$ and $3$ indices are
\ie\label{eqn:m_ind_N23_O(q)}
I^{\mathrm{Sym}^2(T^4)}_\mathrm{NS}=&\frac{2}{q^{\frac12}}+\left(y+\frac{1}{y}\right)-q^{\frac12}\left(6 y^2+12+\frac{6}{y^2}\right)+O\left(q\right)\,,
\\
I^{\mathrm{Sym}^3(T^4)}_\mathrm{NS}=&\frac{3}{q^{\frac34}}+q^{\frac14} \left(y^2+8+\frac{1}{y^2}\right)-8 q^{\frac34} \left(y^3+7 y+\frac{7}{y}+\frac{1}{y^3}\right)+O\left(q^{\frac{5}4}\right)\,.
\fe
We give the expansions of the $N=2$ and $3$ indices up to $q^3$ in Appendix~\ref{app:N=2,3indx}.

\subsection{Exact BPS partition function at $N=2$}

For $N=2$, the index actually contains the information on the exact degeneracy of \( \frac{1}{4} \)-BPS states \cite{Ooguri:1989fd,Benjamin:2021zkn}.
At central charge $c=12$, the contracted large ${\cal N}=4$ superconformal algebra has two short multiplets with $j=0,\,\frac12$ and the NS-sector characters $\chi_{2j}$ and long mulitplets with $j=0$, $h\ge 0$ and the NS-sector character $\chi_{h,j}$. The two short multiplets combine into a long multiplet as
\ie
\chi_0+2\chi_1= \chi_{0,0}\,.
\fe
$\chi_{h,0}$ are related to $\chi_{0,0}$ by $\chi_{h_,0}=q^h \chi_{0,0}$. In Appendix~\ref{app:N=2,3indx}, we give the expansions of the characters $\chi_0$ and $\chi_1$. 

It is reasonable to assume that at a generic point in the moduli space, there are no additional conserved currents \cite{Ooguri:1989fd}.\footnote{We thank Nathan Benjamin for a discussion on this point.}
We can then write down a formula of the partition function of $\frac14$-BPS states (satisfying \eqref{eqn:BPS_cond}),
\ie
Z^\mathrm{BPS}_{N=2}=n_0 \chi_0\overline{\chi_0^\mathrm{BPS}}+n_1\chi_1\overline{\chi_1^\mathrm{BPS}}+\sum_{h=1}^\infty N_h q^h(\chi_0+2\chi_1)\overline{\chi_1^\mathrm{BPS}}\,,
\fe
where $\overline{\chi_{j,{\rm BPS}}}$ consists of the terms in $\overline{\chi_j}$ satisfying the BPS condition, i.e. of the form $\bar y (\bar q^{\frac12}\bar y)^m$ for $m\in\bZ$,
\ie\label{eqn:BPS_chi}
\overline{\chi_0^\mathrm{BPS}}:=\bar q^{\frac32} \bar y^4-2 \bar q \bar y^3+2 \sqrt{\bar q} \bar y^2+\frac{1}{\sqrt{\bar q}}-2\bar y\,,\quad\overline{\chi_1^\mathrm{BPS}}:=-\bar q \bar y^3+2 \sqrt{\bar q} \bar y^2-\bar y\,.
\fe
Taking $\bar y$-derivatives, we find 
\ie
I^{\mathrm{Sym}^2(T^4)}_\mathrm{NS}=\frac12 \bar q^{\frac c{24}-1}\partial_{\bar y}^2 Z^\mathrm{BPS}_{N=2}\big|_{\bar y=\bar q^{-\frac12}}=2n_0\chi_0-n_1\chi_1 - \sum_{h=1}^\infty N_h q^h(\chi_0+2\chi_1)\,.
\fe
The numbers $n_0$, $n_1$ and $N_h$ can be determined by expanding the modified index given in \eqref{eqn:R_modified_index}, \eqref{eqn:NS_modified_index} in terms of the characters. For instance, using the expansions for the index $I^{\mathrm{Sym}^2(T^4)}_\mathrm{NS}$ and the characters $\chi_0$ and $\chi_1$ given in \eqref{eqn:m_ind_N2}, \eqref{eqn:N=2_short_char_j=0}, and \eqref{eqn:N=2_short_char_j=1} in Appendix~\ref{app:N=2,3indx}, we find $n_0=2$, $n_1=5$, $N_2=42$, $N_3=70$, and $N_4=324$. The BPS partition function can be expressed in terms of the modified index as
\ie\label{eqn:BPS_pf}
Z_{{\rm BPS}}^{(N=2)}={\cal S}_0\overline{\chi_0^\mathrm{BPS}}+ {\cal S}_1\overline{\chi_1^\mathrm{BPS}}\,,\quad {\cal S}_0:= \chi_0\,,\quad {\cal S}_1:=2\chi_0-I^{\mathrm{Sym}^2(T^4)}_\mathrm{NS}
\fe
The expansions of ${\cal S}_0$ and ${\cal S}_1$ are
\ie
{\cal S}_0=&\frac{1}{q^{\frac12}}-2\left(y+\frac{1}{y}\right)+q^{\frac12} \left(2 y^2+\frac{2}{y^2}+9\right)-2 q\left(y^3+9 y+\frac{9}{y}+\frac{1}{y^3}\right)+O\left(q^{\frac32}\right)\,,
\\
{\cal S}_1=&-5\left(y+\frac{1}{y}\right)+10 q^{\frac12} \left(y^2+3+\frac{1}{y^2}\right)-5 q \left(y^3+15 y+\frac{15}{y}+\frac{1}{y^3}\right)+O\left(q^{\frac32}\right)\,.
\fe
We give the expansions of ${\cal S}_0=\chi_0$ and ${\cal S}_1$ up to $q^3$ in \eqref{eqn:N=2_short_char_j=0} and \eqref{QuarterBPSDege_2} in Appendix~\ref{app:N=2,3indx}. In Section~\ref{sec:N=2cohomology}, we will compare this partition function with the result from the computation of the $Q$-cohomology.

\section{Lifting as a supercharge cohomology problem}
\label{sec:lifting_as_cohomology}

 In Section~\ref{sec_review_lifting} we reviewed the lifting of $\frac14$-BPS states under second-order conformal perturbation theory about the symmetric-orbifold point. Because the anomalous shifts $\delta \tilde h$ are infinitesimal, the lifting can be analyzed within the free BPS sector, namely the subspace of $\frac14$-BPS states (BPS on the right) at the free orbifold point. At second order, \eqref{anomalous_dimension} reduces, on this sector, to
\ie
\delta \tilde h_2=(\widetilde L_0)_2 = \frac12 \{Q^\dag_1, Q_1\}
\fe
where $Q_1, Q^\dag_1$ are the first-order deformed supercharges defined analogously to \eqref{eqn:n-order-deformed_L-2}. Recognizing the right-hand side as a Laplacian, we may invoke the standard Hodge-theoretic argument to conclude that the subspace of states unlifted at second order is isomorphic to the $Q_1$-cohomology.
This yields a systematic method to identify protected $\frac14$-BPS states via supercharge cohomology.

Since the left-moving generators of the contracted large ${\cal N}=4$ algebra commute with the right-moving supercharge anywhere on the conformal moduli space, recalling as an example the perturbative expansion \eqref{L_-2_expansion} for $L_{-2}$, we have, to the relevant order,
\ie
[(L_{-2})_1,Q_0]+[(L_{-2})_0,Q_1]=0\,.
\fe
It follows that, when acting on the free BPS sector, the zeroth-order left-moving generators map $Q_1$-closed (respectively $Q_1$-exact) states to $Q_1$-closed (respectively $Q_1$-exact) states. Hence, the $Q_1$-cohomology carries a representation of the zeroth-order left-moving contracted large ${\cal N}=4$ superconformal algebra.

In the remainder of this section, we compute the supercharge cohomology at finite $N$. At the orbifold point, operators are labeled by their cycle shapes. In general, states with different cycle shapes may mix under the action of the first-order deformed supercharge. For notational simplicity, we will drop the subscripts and write $Q$ and $Q^\dagger$ for the first-order deformed supercharges.

\subsection{Deformed $Q$-action on the free BPS sector}

\label{sec:Q-action}

Under first-order conformal perturbation theory, the computation of the supercharge $Q$-action involves correlators with a single insertion of the deformation operator \eqref{eqn:exactly_marginal_deformation} which is
in the length-2 single-cycle sector. According to the symmetric group product law, the first-order deformed $Q$-action can change the cycle shapes in two processes: \emph{joining} two cycles into one and \emph{splitting} one cycle into two. More explicitly, the first-order deformed $Q$ decomposes as
\ie\label{eqn:2decom}
Q = Q_{\rm join}+Q_{\rm split}\,.
\fe
Following from the nilpotency of $Q$, the maps $Q_{\rm join}$ and $Q_{\rm split}$ satisfy the nilpotency and mutual anti-commutativity:
\ie\label{eqn:js_comm}
   Q_{\rm join}^2= Q_{\rm split}^2 =\{Q_{\rm join},Q_{\rm split}\} = 0\,.
\fe
In the following, we will loosely refer to the maps $Q_{\rm join}$ and $Q_{\rm split}$ as supercharges.

The supercharges $Q_{\rm join}$ and $Q_{\rm split}$ commute with the diagonal Clifford algebra \eqref{eqn:Clifford}, \eqref{eqn:diag_clif}. Let us depict their actions on a Clifford quartet while keeping track of the right-moving R-charges.
$Q_{\rm join}$ joins two cycles, 
\ie
    \strtwo{w-n}{\tilde j=\frac{w-n-1}{2},\frac{w-n}{2},\frac{w-n+1}{2}} \otimes \strtwo{n}{\tilde j=\frac{n-1}{2},\frac{n}{2},\frac{n+1}{2}} \xlongrightarrow{Q_{\rm join}} \strtwo{w}{\tilde j=\frac{w-1}{2},\frac{w}{2},\frac{w+1}{2}},
\fe
\noindent $Q_{\rm split}$ splits a single cycle into two cycles, 
\ie
    \strtwo{w}{\tilde j=\frac{w-1}{2},\frac{w}{2},\frac{w+1}{2}} \xlongrightarrow{Q_{\rm split}} \strtwo{w-n}{\tilde j=\frac{w-n-1}{2},\frac{w-n}{2},\frac{w-n+1}{2}} \otimes \strtwo{n}{\tilde j=\frac{n-1}{2},\frac{n}{2},\frac{n+1}{2}}.
\fe
The $Q$-action on a general cycle shape is given by a sum over all pairs of cycles, so it suffices to study the $Q$-action on a state with only one or two nontrivial cycles. 

 More explicitly, their actions take the forms
\ie
Q_{\rm join}\ket{\{n_{i_1},(w-n)_{i_2}\}} &= \sum_{j}c_{n_{i_1},(w-n)_{i_2}}^{w_j} \ket{\{w_j\}}\,,
\\
Q_{\rm split}\ket{\{w_i\}} &= \sum_{n=1}^{w-1}\sum_{j_1,j_2}c^{n_{j_1},(w-n)_{j_2}}_{w_i} \ket{\{n_{j_1},(w-n)_{j_2}\}}\,,
\fe
where we write the states using the convention in \eqref{eqn:1/4-BPS_mc}, and for simplicity, we have omitted the $\A,\,\B$ indices of the states, and did not write out the other cycles that do not participate in the process.  We have also assumed the absence of multiplicities, the $N_{m,n}$ in \eqref{eqn:1/4-BPS_mc}. When present, multiplicities appear only in the symmetric factors listed in Table~\ref{tab:covering}. The constraints from the Clifford algebra \eqref{eqn:Clifford} on the $Q$-action will be studied in Section~\ref{sec:N=2cohomology} and Section~\ref{sec:prod_coho}.

By \eqref{eqn:js_comm}, $Q_{\rm join/split}$ can be treated as differentials on a cochain complex, thereby defining a cohomology. In the following, we present our procedure for computing of the $Q_{\rm join/split}$ action, largely borrowing but further extending the technology developed in \cite{Gaberdiel:2023lco}. More precisely, the formulae presented in \cite{Gaberdiel:2023lco} correspond to the $Q_{\rm join}$ process in which one of the cycles being joined is the trivial 1-cycle $1_{--,--}$; this contribution dominates in the large-$N$ limit.

To determine the coefficients $c_{n_{i_1},(w-n)_{i_2}}^{w_j}$ and $c^{n_{j_1},(w-n)_{j_2}}_{w_i}$, we evaluate the matrix elements using conformal perturbation theory. For illustration, we present the procedure to obtain $c_{n_{i_1},(w-n)_{i_2}}^{w_j}$ in detail and briefly outline how to compute the $c^{n_{j_1},(w-n)_{j_2}}_{w_i}$. We have
\ie\label{eqn:cia_computation}
c_{n_{i_1},(w-n)_{i_2}}^{w_j} &= \frac{\bra{\{w_{j}\}} Q \ket{\{n_{i_1},(w-n)_{i_2}\}}}{\bra{\{w_{j}\}}\ket{\{w_{j}\}}} 
\\
&=2 (w-n) n \bra{w_{j}} Q \ket{n_{i_1},(w-n)_{i_2}}
\\
&= 2(w-n) n \oint_{\bar{x}=0} d\bar{x} \int d^2 y \, \bra{w_{j}} \tilde{G}^+(\bar{x}) \Phi(y,\bar{y}) \ket{n_{i_1},(w-n)_{i_2}},
\fe
where \( \tilde{G}^+(\bar{x}) \) is the right-moving supercharge current, and \( \Phi(y,\bar{y}) \) is the exactly marginal deformation operator inserted at position \( (y,\bar{y}) \). From the first to the second line in \eqref{eqn:cia_computation}, we apply the formula \eqref{eqn:SN_symmetrization}, and choose the cycle shapes of the state $\ket{n_{i_1},(w-n)_{i_2}}$ and the supercharge $Q$ to be $(1,2,\cdots,n)(n+1,n+2,\cdots,w)$ and $(n,n+1)$, respectively. 

Following the approach outlined in \cite{Gaberdiel:2023lco}, taking the OPE between the supercurrent and the deformation operator removes the right-moving supercharge $\tilde{G}_{-\frac{1}{2}}$ from the deformation operator. After completing the integrals, the matrix element $\bra{\{w_{j}\}} Q \ket{\{n_{i_1},(w-n)_{i_2}\}}$ is proportional to the three-point function coefficient 
\be
\bra{w_{j}}  V(G_{-\frac{1}{2}}^{'-}\ket{2_{--,--}})(1) \ket{n_{i_1},(w-n)_{i_2}} \label{ThreePT}
\ee
where $V(G_{-\frac{1}{2}}^{'-}\ket{2_{--,--}})(1)$ represents the operator corresponding to the state $G_{-\frac{1}{2}}^{'-}\ket{2_{--,--}}$, inserted at $z=1$. 

The calculation of (\ref{ThreePT}) can be significantly simplified by mapping the computation to the covering space, which we parameterize as the \( t \)-plane, using a covering map
\be
z(t) = \Gamma(t) = t^{w-n} (t-\frac{w}{w-n})^n\,. \label{eqn:largeNcoveringMap}
\ee 
This transformation eliminates the branch cuts in the \( z \)-plane and hence converts the fractional bosonic and fermionic modes into integer or half-integer modes in the covering space. For instance, the fractional modes are mapped as follows:
\ie\label{eqn:frac_mode_transf}
    \alpha_{\frac{m}{w}} &= \oint_{z=0} \partial X^{[m]}(z) \, z^{\frac{m}{w}} \, dz \quad \longrightarrow \quad \oint_{t=0}  \partial X(t) \, \Gamma(t)^{\frac{m}{w}} \, dt\,, 
\\
    \psi_{\frac{m}{w}} &= \oint_{z=0} \psi^{[m]}(z) \, z^{\frac{m}{w}-\frac{1}{2}} \, dz \quad \longrightarrow \quad \oint_{t=0} \psi(t) \left( \frac{d\Gamma(t)}{dt} \right)^{\frac{1}{2}} \Gamma(t)^{\frac{m}{w}-\frac{1}{2}} \, dt\,.
\fe
where $\partial X^{[m]}$ and $\psi^{[m]}$ for $m=0,\,\cdots,\,w-1$ on the $z$-plane are defined in \eqref{eqn:frac_modes} and are lifted to single-valued fields $\partial X$ and $\psi$ on the $t$-plane. Upon performing a series expansion around \( t=0 \), the factors introduced by the covering map \( \Gamma(t) \) yield only integer or half-integer powers of \( t \).

Using the lifting formula \eqref{eqn:frac_mode_transf}, the computation of \eqref{ThreePT} reduces to contour integrals on the $t$-plane, with the integrands being products of the following terms:
\begin{itemize}
    \item Covering map factors: These factors arise from the conformal transformation induced by the covering map \( z = \Gamma(t) \). Their specific form is dictated by the choice of the covering map determined by the cycle shapes of the initial and final states, the left-moving weights of the modes acting on these states, and the fermionic or bosonic nature of the modes. 
    \item Correlation functions: Both the ${\mathbb Z}_2$ twisted operator inserted at \( z = 1 \) and the chiral-chiral primary states involved in $\ket{(w-n)_{i_2}}$ and $\ket{(n)_{i_1}}$ in \eqref{ThreePT} are lifted to spin fields on the covering space (see Appendix~\ref{app: coveringspace}).   Under this map, $\ket{(w-n)_{i_2}}$ is lifted to an operator at $t=0$, while $\ket{(n)_{i_1}}$ is lifted to an operator at $t=\tfrac{w}{\,w-n}$ in the covering space.  Hence, we get a product of a free boson correlator and a correlation function of free fermions with spin fields. The former can be simply computed using Wick contraction, while the latter involving spin fields and fermions can be computed using bosonization techniques with formulae given in the appendix.

\end{itemize}

\begin{table}[t]
    \centering
    \renewcommand{\arraystretch}{1.5}
    \begin{tabular}{|c|c|c|c||c|c|}
        \hline
        supercharge & covering map & Leibniz & conjugate & Leibniz
        \\\hline\hline
        $Q_{\rm join}$ & $t^{w-n} (t-\frac{w}{w-n})^n$ & N & $(Q^\dag)_{\rm split}$ & Y
        \\
        $Q_{\rm split}$ & $t^w (t-\frac{w-n}{w})^{-n}$ & Y & $(Q^\dag)_{\rm join}$ & N 
        \\\hline
    \end{tabular}
    \caption{The covering maps 
    for the supercharges $Q^{(\nu,\ell)}$ and their conjugates, and whether their actions satisfy the Leibniz rule when acting on multi-cycle states. 
    }
    \label{tab:covering}
\end{table}

Finally, let us consider $\widetilde G^-_{\frac12}=Q^\dagger$, the Hermitian conjugate of the first-order deformed conformal supercharge. It admits the decomposition
\ie\label{eqn:3Qdaggers}
Q^\dagger=(Q^\dagger)_{\rm join}+(Q^\dagger)_{\rm split}\,.
\fe
The matrix elements of $(Q^\dagger)_{\rm join}$ and $(Q^\dagger)_{\rm split}$ can be computed following a similar procedure as the one described above for $Q_{\rm join}$. Alternatively, one can use the fact that the first-order deformed $Q^\dagger$ can be obtained from the first-order deformed $Q$ by taking the Hermitian conjugate $\dagger$ in the free orbifold theory. Such a Hermitian conjugate respects the cycle shape; hence, we have 
\ie
(Q^\dagger)_{\rm join}=(Q_{\rm split})^\dagger\,,\quad (Q^\dagger)_{\rm split}=(Q_{\rm join})^\dagger\,.
\fe

The key properties of $(Q^\dagger)_{\rm join}$ and $(Q^\dagger)_{\rm split}$ discussed above are summarized in Table~\ref{tab:covering}.

\subsection{Explicit cohomology representatives at $N=2$}
\label{sec:N=2cohomology}
To illustrate the procedure outlined in the previous subsection, we further study the case of \( N=2 \). There are only two distinct cycle shapes, \((1,1)\) and \((2)\). The $\frac12$-BPS states are
\ie
\ket{\{1_{\pm\pm,\pm\pm},1_{\pm\pm,\pm\pm}\}}\,,\quad \ket{2_{\pm\pm,\pm\pm}}\,,
\fe
and we represent the candidate $\frac14$-BPS states as
\ie
\ket{\{1_{i_1,\pm\pm},1_{i_2,\pm\pm}\}}\,,\quad \ket{2_{j,\pm\pm}}\,,
\fe
where $i$, $i_2$, $j$ denote all possible left-moving modes. 

The supercharge $Q$ can map a state in $V_{(1,1)}$ to $V_{(2)}$ through the joining process, or map a state in $V_{(2)}$ to $V_{(1,1)}$ through splitting. The right-moving Clifford algebra \eqref{eqn:Clifford} imposes selection rules that further constrain the $Q$-action. Since only the diagonal symmetry algebra is preserved under the exactly marginal deformation, we first decompose the states $\ket{\{1_{i_1,\pm\pm},1_{i_2,\pm\pm}\}}$ into four quartets of the diagonal right-moving Clifford algebra, each an eigenvector of the exchange symmetry:
\ie\label{eqn:states_(1,1)S1}
V_{(1,1)S_1}: =\underset{i,j}{\rm span}
\begin{bmatrix}
(\tilde\psi^{[1]+}_{-\frac12}+\tilde\psi^{[2]+}_{-\frac12})(\tilde{\bar\psi}^{[1]+}_{-\frac12}+\tilde{\bar\psi}^{[2]+}_{-\frac12})\ket{\Psi_{(ij)}}
\\
(\tilde{\bar\psi}^{[1]+}_{-\frac 12}+\tilde{\bar\psi}^{[2]+}_{-\frac 12})\ket{\Psi_{(ij)}}
\\
(\tilde\psi^{[1]+}_{-\frac12}+\tilde\psi^{[2]+}_{-\frac12})\ket{\Psi_{(ij)}}
\\
\ket{\Psi_{(ij)}}
\end{bmatrix},
\quad
\tilde j = \begin{bmatrix}
    1 \\ \frac12 \\ \frac12 \\ 0
\end{bmatrix}
\fe
\ie\label{eqn:states_(1,1)S2}
V_{(1,1)S_2}: =\underset{i,j}{\rm span}
\begin{bmatrix}
\tilde\psi^{[1]+}_{-\frac12}\tilde\psi^{[2]+}_{-\frac12}\tilde{\bar\psi}^{[1]+}_{-\frac 12}\tilde{\bar\psi}^{[2]+}_{-\frac 12}\ket{\Psi_{(ij)}}
\\
\tilde\psi^{[1]+}_{-\frac12}\tilde\psi^{[2]+}_{-\frac12}(\tilde{\bar\psi}^{[1]+}_{-\frac 12}-\tilde{\bar\psi}^{[2]+}_{-\frac 12})\ket{\Psi_{(ij)}}
\\
(\tilde\psi^{[1]+}_{-\frac12}-\tilde\psi^{[2]+}_{-\frac12})\tilde{\bar\psi}^{[1]+}_{-\frac 12}\tilde{\bar\psi}^{[2]+}_{-\frac 12}\ket{\Psi_{(ij)}}
\\
(\tilde\psi^{[1]+}_{-\frac12}-\tilde\psi^{[2]+}_{-\frac12})(\tilde{\bar\psi}^{[1]+}_{-\frac 12}-\tilde{\bar\psi}^{[2]+}_{-\frac 12})\ket{\Psi_{(ij)}}
\end{bmatrix},
\quad
\tilde j = \begin{bmatrix}
    2 \\ \frac32 \\ \frac32 \\ 1
\end{bmatrix},
\fe
\ie\label{eqn:states_(1,1)A1}
V_{(1,1)A_1}: =\underset{i,j}{\rm span}
\begin{bmatrix}
    \tilde{\psi}^{[1]+}_{-\frac{1}{2}} \tilde{\psi}^{[2]+}_{-\frac{1}{2}}(\tilde{\bar\psi}^{[1]+}_{-\frac 12}+\tilde{\bar\psi}^{[2]+}_{-\frac 12})\ket{\Psi_{[ij]}} 
    \\
    (\tilde{\psi}^{[1]+}_{-\frac{1}{2}}-\tilde{\psi}^{[2]+}_{-\frac{1}{2}}) 
    (\tilde{\bar\psi}^{[1]+}_{-\frac 12}+\tilde{\bar\psi}^{[2]+}_{-\frac 12})\ket{\Psi_{[ij]}} 
    \\
    \tilde{\psi}^{[1]+}_{-\frac{1}{2}} \tilde{\psi}^{[2]+}_{-\frac{1}{2}} \ket{\Psi_{[ij]}} 
    \\
    (\tilde{\psi}^{[1]+}_{-\frac{1}{2}}-\tilde{\psi}^{[2]+}_{-\frac{1}{2}}) \ket{\Psi_{[ij]}},
\end{bmatrix},
\quad
\tilde j = \begin{bmatrix}
    \frac32 \\ 1 \\ 1 \\ \frac12
\end{bmatrix}
\fe
\ie\label{eqn:states_(1,1)A2}
V_{(1,1)A_2}: =\underset{i,j}{\rm span}
\begin{bmatrix}
    \tilde{\bar\psi}^{[1]+}_{-\frac{1}{2}} \tilde{\bar\psi}^{[2]+}_{-\frac{1}{2}}(\tilde{\psi}^{[1]+}_{-\frac 12}+\tilde{\psi}^{[2]+}_{-\frac 12})\ket{\Psi_{[ij]}} 
    \\
    (\tilde{\bar\psi}^{[1]+}_{-\frac{1}{2}}-\tilde{\bar\psi}^{[2]+}_{-\frac{1}{2}}) 
    (\tilde{\psi}^{[1]+}_{-\frac 12}+\tilde{\psi}^{[2]+}_{-\frac 12})\ket{\Psi_{[ij]}} 
    \\
    \tilde{\bar\psi}^{[1]+}_{-\frac{1}{2}} \tilde{\bar\psi}^{[2]+}_{-\frac{1}{2}} \ket{\Psi_{[ij]}} 
    \\
    (\tilde{\bar\psi}^{[1]+}_{-\frac{1}{2}}-\tilde{\bar\psi}^{[2]+}_{-\frac{1}{2}}) \ket{\Psi_{[ij]}} 
\end{bmatrix},
\quad
\tilde j = \begin{bmatrix}
    \frac32 \\ 1 \\ 1 \\ \frac12
\end{bmatrix},
\fe
where $\ket{\Psi_{ij}} := |1_{i,--},1_{j,--}\rangle$ and $(ij)$, $[ij]$ denotes symmetrization and antisymmetrization. Similarly, the space $V_{(2)}$ is spanned by
\ie\label{eqn:V(2)}
V_{(2)}: =\underset{k}{\rm span}
\begin{bmatrix}
(\tilde\psi^{[1]+}_{-\frac12}+\tilde\psi^{[2]+}_{-\frac12})(\tilde{\bar\psi}^{[1]+}_{-\frac12}+\tilde{\bar\psi}^{[2]+}_{-\frac12})\ket{\Psi_{k}}
\\
(\tilde{\bar\psi}^{[1]+}_{-\frac 12}+\tilde{\bar\psi}^{[2]+}_{-\frac 12})\ket{\Psi_{k}}
\\
(\tilde\psi^{[1]+}_{-\frac12}+\tilde\psi^{[2]+}_{-\frac12})\ket{\Psi_{k}}
\\
\ket{\Psi_{k}}
\end{bmatrix},
\quad
\tilde j = \begin{bmatrix}
    \frac32 \\ 1 \\ 1 \\ \frac12
\end{bmatrix}\,,
\fe
where $\ket{\Psi_{k}} := |2_{k,--}\rangle$.

Recall the following facts:
\begin{enumerate}
    \item The supercharge \( Q \) carries right-moving R-charge $\tilde j=\frac12$;
    \item \( Q \) commutes with the right-moving Clifford algebra in the free BPS sector;
\end{enumerate}
Now, consider $Q$ acting on states in $V_{(1,1)A_2}$, which must give a state in $V_{(2)}$. Since $Q$ acting on the top component gives a state with $\tilde j = \frac32 + \frac12 = 2$ exceeding the $\tilde j$ of $V_{(2)}$ given in \eqref{eqn:V(2)}, the $Q$-action must annihilate the states in $V_{(1,1)A_2}$. Repeatedly applying similar arguments, we arrive at
the following cochain complexes:
\ie
0\overset{Q}{\longrightarrow}
V_{(1,1)S_1}
&\overset{Q}{\longrightarrow}
V_{(2)}
\overset{Q}{\longrightarrow}
V_{(1,1)S_2}
\overset{Q}{\longrightarrow} 0\,,
\\
0 & \overset{Q}{\longrightarrow} V_{(1,1)A_1}\overset{Q}{\longrightarrow} 0\,,
\\
0 & \overset{Q}{\longrightarrow} V_{(1,1)A_2}\overset{Q}{\longrightarrow} 0\,. \label{eqn:N=2complex}
\fe
We observe that the states in \(V_{(1,1)A_1}\) and \(V_{(1,1)A_2}\) must represent non-trivial supercharge cohomology classes. 

Applying similar arguments to the $Q^\dagger$-action, we obtain the following complexes:
\ie
0\overset{Q^\dagger}{\longrightarrow} 
V_{(1,1)S_2}
&\overset{Q^\dagger}{\longrightarrow}
V_{(2)}
\overset{Q^\dagger}{\longrightarrow} 
V_{(1,1)S_1}
\overset{Q^\dagger}{\longrightarrow}  0\,,
\\
0 & \overset{Q^\dagger}{\longrightarrow}  V_{(1,1)A_1}\overset{Q^\dagger}{\longrightarrow}  0\,,
\\
0 & \overset{Q^\dagger}{\longrightarrow}  V_{(1,1)A_2}\overset{Q^\dagger}{\longrightarrow}  0\,. \label{eqn:N=2complex_dagger}
\fe
Since the states in \( V_{(1,1)A_1} \) and \( V_{(1,1)A_2} \) are annihilated by both \( Q \) and \( Q^\dagger \), they are BPS \emph{states} and remain unlifted under the exactly marginal deformation. In addition to these, there are also BPS states in \( V_{(2)} \), which correspond in one-to-one to the cohomology classes of cycle shape $(2)$.

The dimensions of the cohomology groups give the degeneracy of \( \frac{1}{4} \)-BPS states, which could be summarized into a BPS partition function, where the right-moving part of the states in the $V_{(1,1)S_i}$, $V_{(1,1)A_i}$, and $V_{(2)}$ contribute $\overline{\chi_0^\mathrm{BPS}}$, $2\overline{\chi_1^\mathrm{BPS}}$, and $\overline{\chi_1^\mathrm{BPS}}$ given in \eqref{eqn:BPS_chi}, respectively. We computed the Q cohomology explicitly using the method in subsection \ref{sec:Q-action} up to $h=4$ . We compare the degeneracies of Q Cohomology classes with the expression in \eqref{eqn:BPS_pf}, using the expansions of \({\cal S}_0\) and \({\cal S}_1\) in \eqref{eqn:N=2_short_char_j=0} and \eqref{QuarterBPSDege_2}, and find an exact match.

In the following discussion, we omit the right-moving structure, as the Clifford algebra relates the different components of the right-moving quartet. We present the representatives for the full cohomology classes in the $h=1,j=0$ sector here and in the $h=\frac{3}{2}, j=\frac{1}{2}$ sector in appendix \ref{sec:Cohomology_h32}.  

In the twisted (2) sector, the representatives are
\bea
&&\bar\psi^+_{-\frac{1}{2}}\psi^-_0\bar\psi^-_0\ket{2_{--}},\quad \psi^-_0\psi^+_{-\frac{1}{2}}\bar\psi^-_0\ket{2_{--}}\crcr
&&(\psi^-_0\bar\alpha^2_{-\frac{1}{2}}+\bar\psi^-_0\alpha^2_{-\frac{1}{2}})\ket{2_{--}}\crcr
&&\bar\psi^-_0 \alpha^1_{-\frac{1}{2}}\ket{2_{--}},\quad \bar\psi^-_{-\frac{1}{2}}\ket{2_{--}}, \psi^-_{-\frac{1}{2}}\ket{2_{--}}
\eea
In the untwisted $(1,1)_{S_1}$ sector, the representatives are
\bea
&& (-\bar\psi^{[i]+}_{-\frac{1}{2}}\psi^{[i]-}_{-\frac{1}{2}}+\bar\psi^{[j]-}_{-\frac{1}{2}}\psi^{[i]+}_{-\frac{1}{2}}) \ket{1_{--},1_{--}},\quad   (-\bar\psi^{[i]+}_{-\frac{1}{2}}\bar\psi^{[i]-}_{-\frac{1}{2}}+\bar\psi^{[j]-}_{-\frac{1}{2}}\bar\psi^{[i]+}_{-\frac{1}{2}})  \ket{1_{--},1_{--}}\crcr
&&(\psi^{[i]-}_{-\frac{1}{2}} \psi^{[i]+}_{-\frac{1}{2}} -\psi^{[i]-}_{-\frac{1}{2}} \psi^{[j]+}_{-\frac{1}{2}}) \ket{1_{--},1_{--}},\quad (-\bar\psi^{[i]+}_{-\frac{1}{2}}\psi^{[i]-}_{-\frac{1}{2}}+\psi^{[j]-}_{-\frac{1}{2}}\bar\psi^{[i]+}_{-\frac{1}{2}})\ket{1_{--},1_{--}},\crcr
&&(-\bar\psi^{[i]+}_{-\frac{1}{2}}\psi^{[i]-}_{-\frac{1}{2}}+\psi^{[i]+}_{-\frac{1}{2}}\bar\psi^{[i]-}_{-\frac{1}{2}} )\ket{1_{--},1_{--}}\crcr
&&\bar\alpha^{[i]2}_{-1}\ket{1_{--},1_{--}},\quad \alpha^{[i]1}_{-1} \ket{1_{--},1_{--}},\quad \bar\alpha^{[i]1}_{-1} \ket{1_{--},1_{--}},\quad \alpha^{[i]2}_{-1} \ket{1_{--},1_{--}}\label{eqn:11sBPSstatesh1j0}
\eea
where the summation of $i,j$ ranging from 1 to 2 is implicit to ensure orbifold invariance. 
In the $(1,1)_A$ sector, the representatives are
\bea
&& (\alpha^{[1]1}_{-1}-\alpha^{[2]1}_{-1})\ket{1_{--},1_{--}}, \quad (\bar\alpha^{[1]1}_{-1}-\bar\alpha^{[2]1}_{-1})\ket{1_{--},1_{--}},\quad (\bar\alpha^{[1]2}_{-1}-\bar\alpha^{[2]2}_{-1})\ket{1_{--},1_{--}}\crcr
&&(\alpha^{[1]2}_{-1}-\alpha^{[2]2}_{-1})\ket{1_{--},1_{--}}, \quad (\psi^{[1]-}_{-\frac{1}{2}}\psi^{[1]+}_{-\frac{1}{2}}-\psi^{[2]-}_{-\frac{1}{2}}\psi^{[2]+}_{-\frac{1}{2}}) \ket{1_{--},1_{--}},\quad (\bar\psi^{[1]+}_{-\frac{1}{2}}\psi^{[1]-}_{-\frac{1}{2}}-\bar\psi^{[2]+}_{-\frac{1}{2}}\psi^{[2]-}_{-\frac{1}{2}}) \ket{1_{--},1_{--}}\crcr
&&(\bar\psi^{[1]+}_{-\frac{1}{2}}\bar\psi^{[1]-}_{-\frac{1}{2}}-\bar\psi^{[2]+}_{-\frac{1}{2}}\bar\psi^{[2]-}_{-\frac{1}{2}}) \ket{1_{--},1_{--}},\quad (\psi^{[1]+}_{-\frac{1}{2}}\bar\psi^{[1]-}_{-\frac{1}{2}}-\psi^{[2]+}_{-\frac{1}{2}}\bar\psi^{[2]-}_{-\frac{1}{2}}) \ket{1_{--},1_{--}}, \crcr
&&(\psi^{[2]-}_{-\frac{1}{2}}\bar\psi^{[1]+}_{-\frac{1}{2}}-\psi^{[1]-}_{-\frac{1}{2}}\bar\psi^{[2]+}_{-\frac{1}{2}})\ket{1_{--},1_{--}},\quad (\psi^{[2]-}_{-\frac{1}{2}}\psi^{[1]+}_{-\frac{1}{2}}-\psi^{[1]-}_{-\frac{1}{2}}\psi^{[2]+}_{-\frac{1}{2}})\ket{1_{--},1_{--}}\crcr
&&(\bar\psi^{[2]-}_{-\frac{1}{2}}\bar\psi^{[1]+}_{-\frac{1}{2}}-\bar\psi^{[1]-}_{-\frac{1}{2}}\bar\psi^{[2]+}_{-\frac{1}{2}})\ket{1_{--},1_{--}},\quad (\bar\psi^{[2]-}_{-\frac{1}{2}}\psi^{[1]+}_{-\frac{1}{2}}-\bar\psi^{[1]-}_{-\frac{1}{2}}\psi^{[2]+}_{-\frac{1}{2}})\ket{1_{--},1_{--}}
\eea
where the antisymmetrization of the right-moving sector is implicit.

\section{Monotone and fortuitous cohomologies}

\subsection{Classification}
\label{sec:defns}
A classification scheme of supercharge $Q$-cohomology classes was proposed in \cite{Chang:2024zqi} based on their properties at large $N$. The classification relies on a property that the vector space of operators (physical Hilbert space, but without the norm) in the finite $N$ theory can be expressed as a quotient of that in the infinite $N$ theory by certain equivalent relations. A monotone cohomology class has a representative that can be pulled back to a representative of a nontrivial cohomology class in the infinite $N$ theory. By contrast, the pullback of a representative of a fortuitous cohomology class is not $Q$-closed.

In the D1-D5 CFT, the quotient map is known in the literature as the ``stringy exclusion principle'' (SEP) \cite{Maldacena:1998bw}. However, as we will see, the SEP map does not commute with the supercharge, so the classification proposed in \cite{Chang:2024zqi} cannot be directly applied to the D1-D5 CFT. Instead, we require a modification the original classification scheme of \cite{Chang:2024zqi}; this modification is similar to that introduced in \cite{Chang:2024lxt}.

Let us begin by giving a precise definition of the SEP map. Consider a D1-D5 CFT of rank $N'$, greater than the rank $N$ of the D1-D5 CFT we are interested in.  Both CFTs are at their free orbifold points. The SEP map $\pi$ is a map between the free BPS sectors
\ie\label{eqn:SEP_map}
\pi_{N,N'}:~V_{N'}\to V_{N}\,,\quad N'>N
\fe
to be defined below. 

We will first consider a map
\ie\label{eqn:SEP_map}
\pi^{\rm pre}_{N,N'}:~V_{N'}\to V_{N}\,,\quad N'>N\,,
\fe
discuss its disadvantages, and promote it to the map $\pi_{N,N'}$. As we have seen in \eqref{eqn:Hp}, \eqref{eqn:full_Hilbert_space} and Table~\ref{tab:H}, the Hilbert space $\cH_N$ of the rank $N$ theory is given by the $S_N$ projection of the direct sum of the Hilbert spaces $\cH_g$ (in the product theory before $S_N$ projection) twisted by the group elements $g\in S_N$. This property also holds for the top spaces for BPS sectors $V_{N'}$ and $V_{N}$. Hence, we can define a map $\pi^{\rm pre}_{N,N'}$ via a collection of maps 
\ie
\{ (\pi^{\rm pre}_{N,N'})^{g, g'} :~V_{g'}\to V_{g} \mid g'\in S_{N'}~{\rm and}~g\in S_N \},
\fe
which are specified as follows.
For $g'\neq g(N+1)(N+2)\cdots(N')$, the map $(\pi^{\rm pre}_{N,N'})^{g, g'}$ projects out all the states
\ie
(\pi^{\rm pre}_{N,N'})^{g, g'} V_{g'} = 0\,.
\fe
For $g'= g(N+1)(N+2)\cdots(N')$, we have
\ie
V_{g'} = \left(V_g\otimes V_{(1)}^{\otimes(N'-N)}\right)\,,
\fe
and the map $(\pi^{\rm pre}_{N,N'})^{g, g'}$ acts on the factor $V_g$ as the identity map and on each of the factor $V_{(1)}$'s as a projection onto the vacuum state $\ket 1:=\ket{1_{--,--}}$.\footnote{We have implicitly used the trivial isomorphism $\cH_g\otimes\ket 1^{\otimes(N'-N)}\cong \cH_g$.} In other words, $\pi^{\rm pre}_{N,N'}$ satisfies
\ie
(\pi^{\rm pre}_{N,N'})^{g, g'} = 1_{V_g}\otimes \big(\ket 1 \bra 1\big)^{\otimes (N'-N)}\,.
\fe

Due to the choice of $\ket{1}$ above, the map $\pi^{\rm pre}_{N,N'}$ commutes with the ``$+$" half of the right-moving diagonal Clifford algebra but does not commute with $\tilde\psi^{\rm diag-}_{\frac12}$ and $\bar{\tilde\psi}^{\rm diag-}_{\frac12}$, i.e.\footnote{Note that $\pi^{\rm pre}_{N,N'}$ acts on the right-moving space as
\ie
\pi^{\rm pre}_{N,N'}\big|_{\text{right-moving}}=\prod_{i=N+1}^{N'}\tilde\psi^{[i]-}_{\frac12}\tilde\psi^{[i]+}_{-\frac12}\bar{\tilde\psi}^{[i]-}_{\frac12}\bar{\tilde\psi}^{[i]+}_{-\frac12}\,.
\fe}
\ie
\pi^{\rm pre}_{N,N'}\tilde\psi^{\rm diag+}_{-\frac12}=\tilde\psi^{\rm diag+}_{-\frac12}\pi^{\rm pre}_{N,N'}\,,\quad \pi^{\rm pre}_{N,N'}
\bar{\tilde\psi}^{\rm diag+}_{-\frac12}=\bar{\tilde\psi}^{\rm diag+}_{-\frac12}\pi^{\rm pre}_{N,N'}\,,
\fe
, where the $\tilde\psi^{\rm diag+}_{-\frac12}$ and  $\bar{\tilde\psi}^{\rm diag+}_{-\frac12}$ appearing to the right of $\pi^{\rm pre}_{N,N'}$ are defined in the
theory with $N$, whereas those appearing to its left are defined in the
theory with $N'$. The same convention applies to
Eqs.~(\ref{eqn: piNNp1}) and (\ref{eqn: piNNp2}). Since the right-moving diagonal Clifford algebra plays an important rule in our analysis, we seek an SEP map $\pi_{N,N'}$ that fully commutes with it.
Define
\ie
\pi_{N,N'} := ~ &\tilde\psi^{\rm diag-}_\frac12\bar{\tilde\psi}^{\rm diag-}_\frac12\pi_{N,N'}^{\rm pre}\tilde\psi^{\rm diag+}_{-\frac12}\bar{\tilde\psi}^{\rm diag+}_{-\frac12}+\tilde\psi^{\rm diag+}_{-\frac12}\bar{\tilde\psi}^{\rm diag-}_\frac12\pi_{N,N'}^{\rm pre}\tilde\psi^{\rm diag-}_\frac12\bar{\tilde\psi}^{\rm diag+}_{-\frac12}
\\
&+\tilde\psi^{\rm diag-}_\frac12\bar{\tilde\psi}^{\rm diag+}_{-\frac12}\pi_{N,N'}^{\rm pre}\tilde\psi^{\rm diag+}_{-\frac12}\bar{\tilde\psi}^{\rm diag-}_\frac12+\tilde\psi^{\rm diag+}_{-\frac12}\bar{\tilde\psi}^{\rm diag+}_{-\frac12}\pi_{N,N'}^{\rm pre}\tilde\psi^{\rm diag-}_\frac12\bar{\tilde\psi}^{\rm diag-}_\frac12\,. \label{eqn: piNNp1}
\fe
A straightforward computation then shows
\ie
&\pi_{N,N'}\tilde\psi^{\rm diag+}_{-\frac12}=\tilde\psi^{\rm diag+}_{-\frac12}\pi_{N,N'}\,,\quad \pi_{N,N'}
\bar{\tilde\psi}^{\rm diag+}_{-\frac12}=\bar{\tilde\psi}^{\rm diag+}_{-\frac12}\pi_{N,N'}\,,
\\
&\pi_{N,N'}\tilde\psi^{\rm diag-}_{\frac12}=\tilde\psi^{\rm diag-}_{\frac12}\pi_{N,N'}\,,\quad \pi_{N,N'}
\bar{\tilde\psi}^{\rm diag-}_{\frac12}=\bar{\tilde\psi}^{\rm diag-}_{\frac12}\pi_{N,N'}\,.\label{eqn: piNNp2}
\fe
We also note that the SEP maps compose nicely as 
\ie\label{eqn:pi_comp}
\pi_{N_1,N_3}=\pi_{N_1,N_2}\circ\pi_{N_2,N_3}:~V_{N_3}\to V_{N_2}\to V_{1}\,,\quad N_3>N_2>N_1\,.
\fe

After defining the SEP map $\pi_{N,N'}$, we study the interplay between $\pi_{N,N'}$ and the first-order defromed $Q$. To make our discussion more precise, we write $Q_N$ and $Q_{N'}$ as the supercharges acting on $V_N$ and $V_{N'}$, respectively. Consider a nontrivial $Q_N$-cohomology class represented by a $Q_N$-closed state $\ket{\Psi}\in V_N$, which can be further lifted to a state $\ket{\Psi'}$ in $V_{N'}$ by the SEP map $\pi_{N,N'}$, i.e.
\ie
\ket{\Psi'}\in\pi_{N,N'}^{-1}(\ket{\Psi})\subset V_{N'}\,.
\fe
The state \( \ket{\Psi'} \) must fall into one of the following three scenarios:
\begin{enumerate}
\item \( \ket{\Psi'} \) is \( Q_{N'} \)-closed but not \( Q_{N'} \)-exact. 
\item \( \ket{\Psi'} \) is not \( Q_{N'} \)-closed. 
\item \( \ket{\Psi'} \) is \( Q_{N'} \)-exact. 
\end{enumerate}
In the first scenario, the $Q_{N'}$-closed state represents a nontrivial $Q_{N'}$-cohomology class and can be promoted---by adding suitable \( Q_{N'} \)-exact states---to a BPS state in \( V_{N'} \). In both the second and third cases, \( \ket{\Psi'} \) is irredeemably non-BPS and cannot be promoted to a BPS state.

The third scenario is forbidden if the SEP map $\pi_{N,N'}$ commutes with the supercharge; more precisely
\ie\label{eqn:pi_Q_commute}
\pi_{N,N'} Q_{N'}=Q_N\pi_{N,N'}\,,
\fe
since $\ket{\Psi} = \pi_{N,N'}\ket{\Psi'}=\pi_{N,N'} Q_{N'}\ket{\chi}= Q_N\pi_{N,N'}\ket{\chi}$, contradicting the assumption that $\ket{\Psi}$ represents a nontrivial class. However, unlike in the ${\cal N}=4$ SYM \cite{Chang:2024zqi}, in the D1-D5 CFTs, the SEP map $\pi_{N,N'}$ does not always commute with the supercharge. As an illustrative example, let's consider the $Q, \pi_{2,3}$ commutator acting on a particular two-cycle state with cycle shape $(1,2)$ in $N=3$,
\ie
&(\pi_{2,3} Q_{3}-Q_{2}\pi_{2,3})\left[\psi^{[1]-}_{-\frac{1}{2}} \ket{1_{--,++}} \otimes\alpha^2_{-\frac{1}{2}}\alpha^2_{-\frac{1}{2}}\ket{2_{--,--}}+(S_3~{\rm permutations})\right]
\\
&\propto\psi^{[1]-}_{-\frac{1}{2}} \psi^{[2]+}_{-\frac{1}{2}}\alpha^{[2]2}_{-1} \tilde{\psi}^{[2]+}_{-\frac{1}{2}}\tilde{\bar\psi}^{[2]+}_{-\frac{1}{2}}\ket{1_{--,+ +},1_{--,--}} +(S_2~{\rm permutations})\,.
\fe

We now define the monotone and fortuitous cohomology classes. Set
\ie
U^{\rm closed}_N&:=\big\{|\Psi\rangle\in {\rm Ker}(Q_N) \,\big|\, {\rm Ker}(Q_{N'})\cap\pi_{N,N'}^{-1}(|\Psi\rangle)\neq 0~~\text{for all } N' > N ~~{\rm or}~~ |\Psi\rangle = 0\big\}\,,
\\
U^{\rm exact}_N&:=\big\{|\Psi\rangle\in {\rm Ker}(Q_N) \,\big|\, {\rm Im}(Q_{N'})\cap\pi_{N,N'}^{-1}(|\Psi\rangle)\neq 0 ~~\text{for any } N' > N ~~{\rm or}~~ |\Psi\rangle = 0\big\}\,,
\fe
which are subspaces of $Q_N$-closed states that either can lift via $\pi_{N,N'}$ to $Q_{N'}$-closed states or to $Q_{N'}$-exact states, respectively (or are zero). To see $U^{\rm closed}_N$ and $U^{\rm exact}_N$ are vector spaces, let us rewrite them as
\ie
U^{\rm closed}_N &= {\rm Ker}(Q_N)\cap\,\bigcap_{N'}\,{\rm Im}\left(\pi_{N,N'}\big|_{{\rm Ker}(Q_{N'})}\right)\,,
\\
U^{\rm exact}_N &={\rm Ker}(Q_N)\cap \sum_{N'}{\rm Im}(\pi_{N,N'} Q_{N'})\,.
\fe
We define the spaces of monotone and fortuitous cohomology classes as follows.
\begin{defn}[Monotone and fortuitous]\label{def:monotone}
The space of monotone classes is
\ie\label{eqn:defmon}
\cH^{\rm mon}_N:=(U^{\rm closed}_N+{\rm Im}(Q_N))/{\rm Im}(Q_N)\,.
\fe
The space of fortuitous classes is given by the quotient of the $Q$-cohomology $H^*_Q(V_N)$ by the space of monotone classes 
\ie
\cH^{\rm for}_N:=H^*_Q(V_N)/\cH^{\rm mon}_N\cong {\rm Ker}(Q_N)\big/\big(U^{\rm closed}_N+{\rm Im}(Q_N)\big)\,.
\fe
\end{defn}
\noindent In other words, monotone classes of rank $N$ can lift to monotone $Q_{N'}$-cohomology classes, whereas fortuitous classes cannot. By the third scenario, a monotone class that is inside a subspace $\cH^{\rm trivial}_N\subset \cH^{\rm mon}_N$ may lift to a trivial class at rank $N'$, where $\cH^{\rm trivial}_N$ is defined by
\ie
\cH^{\rm trivial}_N:=(U^{\rm exact}_N+{\rm Im}(Q_N))/{\rm Im}(Q_N)\,.
\fe
To account for this, we define the absolute monotone class as follows.
\begin{defn}[Absolute monotone]\label{def:largeNexact}
The space of absolute monotone classes is
\ie
\cH^{\rm abs\,mon}_N:=\cH^{\rm mon}_N/\cH^{\rm trivial}_N\cong \big(U^{\rm closed}_N+{\rm Im}(Q_N)\big)\big/\big(U^{\rm exact}_N+{\rm Im}(Q_N)\big)\,.
\fe
\end{defn}
\noindent In other words, an absolute monotone class must lift to non-trivial cohomology classes at higher ranks.

It is easy to see that when the SEP map commutes with the supercharge, i.e. \eqref{eqn:pi_Q_commute}, we have $\cH^{\rm trivial}_N=0$. Hence, a non-trivial monotone class at rank $N$ must lift to non-trivial monotone classes at higher rank $N'$. This property can be seen more explicitly as follows. By \eqref{eqn:pi_Q_commute} and the composition rule \eqref{eqn:pi_comp}, the SEP map acts between the spaces $U^{\rm closed}_N$ as a projection,
\ie
\pi_{N,\widetilde N}:~U^{\rm closed}_{\widetilde N}\to U^{\rm closed}_{ N}\,,\quad \widetilde N>N\,.
\fe
Such a projection induces a projection map $(\pi_{N,\widetilde N})_*$ on the space of monotone classes,
\ie
(\pi_{N,\widetilde N})_*:~\cH^{\rm mon}_{\widetilde N}\to \cH^{\rm mon}_N\,,
\fe
which implements the aforementioned lifting between non-trivial monotone classes. However, since \eqref{eqn:pi_Q_commute} does not always hold in the D1-D5 CFTs, the map $(\pi_{N,\widetilde N})_*$ does not always exist.

From Definition~\ref{def:monotone}, deciding whether a rank-$N$ cohomology class is monotone or fortuitous is generally infeasible: constructing $U^{\rm closed}_N$ requires knowing the $Q_{N'}$-closed subspaces for arbitrarily large $N'>N$. In practice, we therefore work with the finite-cutoff proxy $\cH^{\rm mon_{N'}}_N$, defined for $N'>N$ by
\ie
\cH_N^{\rm mon_{N'}}&=(U^{\rm closed_{N'}}_N+{\rm Im}(Q_N))/{\rm Im}(Q_N)\,,
\\
U^{\rm closed_{N'}}_N &= {\rm Ker}(Q_N)\cap{\rm Im}\left(\pi_{N,N'}\big|_{{\rm Ker}(Q_{N'})}\right)\,.
\fe
For later discussions, we also define
\ie
\cH_N^{{\rm trivial}_{N'}}&=(U^{{\rm exact}_{N'}}_N+{\rm Im}(Q_N))/{\rm Im}(Q_N)\,,
\\
U^{{\rm exact}_{N'}}_N &= {\rm Ker}(Q_N)\cap{\rm Im}\left(\pi_{N,N'}Q_{N'}\right)\,.
\fe
It is easy to see that $\cH_N^{{\rm mon}_{N'}}\supset \cH_N^{\rm mon}$; hence, $\cH_N^{{\rm mon}_{N'}}$ contains all the monotone classes, but it may also contain some fortuitous classes. However, we believe that $\cH_N^{\rm mon_{N'}}$ is a good approximation of $\cH^{\rm mon}_N$ in most of the cases.

In Section~\ref{sec:N=1classification}, we study the space $\cH_1^{\rm mon_2}$. We leave the study of $\cH_1^{\rm mon_3}$ and $\cH_2^{\rm mon_3}$ to the follow-up paper \cite{Chang:2025xx}.

\subsection{Example: classification at $N=1$}
\label{sec:N=1classification}

The SEP map $\pi_{1,2}$ acts between the $N'=2$ and $N=1$ cochain complexes as
\ie
\begin{tikzcd}
 0 \arrow[r]  & V_{(1,1)S_1} \arrow[r,"Q_2"] \arrow[d,"\pi_{1,2}"] & V_{(2)} \arrow[r] \arrow[d,"\pi_{1,2}"] &\cdots
\\
0 \arrow[r] & V_{(1)} \arrow[r,"Q_1"] & 0 
\end{tikzcd}
\fe
It follows that $\pi_{1,2}$ commutes with the supercharge, i.e.
\ie
\pi_{1,2} Q_2=Q_1\pi_{1,2}\,.
\fe
Hence, nontrivial $N=1$ classes cannot lift to a trivial $N=2$ class, i.e. $\cH^{\rm trivial_2}_1=0$. 

In the $N=1$ theory, all $\frac14$-BPS states are the contracted large ${\cal N}=(4,4)$ descendants of the half-BPS quartets. These states can only lift to the $(1,1)_{S_1}$ sector at $N=2$, because the R-charge $\tilde j$ and the diagonal Clifford algebra are preserved by the SEP map. We have determined the lifts of $N=1$ BPS states to $N=2$ explicitly up to $h=3$. Following the definition \ref{def:monotone}, our result suggests that all $N=1$ cohomology classes are monotones. 

\subsubsection{Monotone stringy exclusion principle}
We find the $\mathcal{H}_1^{\textrm{mon}2}$ monotone partition function at $N=1$ is given by 
\be
Z_{N=1}^{\textrm{mon}2} = 1 + 2 q^{\frac{1}{2}}(y+y^{-1})+q(8+2y^{-2}+2y^2) +\cdots
\ee
In general, the number of $N=2$ $Q$-closed states that map to $N=1$ BPS states under $\pi_{1,2}$ exceeds the actual degeneracy of the $N=1$ states.

As a demonstration, there are 8 monotone states at $h=1,j=0$ in $N=1$: 
\bea
&&\alpha^{1}_{-1}\ket{0}, \bar\alpha^{1}_{-1}\ket{0},
\alpha^{2}_{-1}\ket{0},
\bar\alpha^{2}_{-1}\ket{0}\crcr
&& \psi^-_{-\frac{1}{2}}\psi^+_{-\frac{1}{2}}\ket{0}, \bar\psi^+_{-\frac{1}{2}}\psi^-_{-\frac{1}{2}}\ket{0}, \bar\psi^+_{-\frac{1}{2}}\bar\psi^-_{-\frac{1}{2}}\ket{0}, \psi^+_{-\frac{1}{2}}\bar\psi^-_{-\frac{1}{2}}\ket{0}\label{eqn:N1BPS}
\eea
We found nine cohomology representatives in the $(1,1)_{S_1}$ sector, as listed in (\ref{eqn:11sBPSstatesh1j0}). All of them can be projected to states in \eqref{eqn:N1BPS}. However, the projections of the first and fourth states are related through the projection of the fifth state. Consequently, although we begin with 9 supercharge cohomology representatives at $N=2$, only 8 independent cohomology representatives remain after imposing the SEP map.

We expect this phenomenon to happen in general, and we call it the monotone stringy exclusion principle, where independent cohomology representatives at larger $N$ are related to each other at smaller $N$ after imposing the SEP map. \footnote{We thank Ji Hoon Lee for insightful discussions on this point.}

\section{Composite BPS states}
\label{sec:composite_BPS}
This section uses single-cycle BPS states to construct composite BPS states. We depart from the $Q$-cohomology framework and consider actual BPS \emph{states} because analyzing the $Q^\dag$ action is simpler than establishing $Q$-non-exactness.

\subsection{When is a two-cycle state BPS?}
\label{sec:prod_coho}

Let us consider two-cycle states in $\mathrm{Sym}^N(T^4)$\footnote{At this point, we do not impose any further assumptions (e.g. closedness).}
\ie\label{eqn:product_state}
\ket{\{(w_1)_{i,\pm\pm},(w_2)_{j,\pm\pm}\}}\,,\quad w_1+w_2=N\,.
\fe
As we have seen in Section~\ref{sec:N=2cohomology}, the right-moving Clifford algebra imposes selection rules on the $Q$-action, which could forbid certain joining and splitting processes. Let us perform similar analysis on our setup here.

The Clifford algebra generators of the length-$w_1$ and length-$w_2$ cycles are
\ie
\tilde{\psi}^{(w_1)\pm}_{-\frac 12}\,,\quad \tilde{\bar\psi}^{(w_1)\pm}_{-\frac 12}\,,\quad\tilde{\psi}^{(w_2)\pm}_{-\frac 12}\,,\quad \tilde{\bar\psi}^{(w_2)\pm}_{-\frac 12}\,,
\fe
where the definition of $\psi_{r+\frac mw}^{(w)}$ are given in \eqref{eqn:frac_modes}. The diagonal Clifford algebra generators are
\ie
\tilde{\psi}^{{\rm diag}\pm}_{-\frac 12}=\sqrt{\frac{w_1}{w_1+w_2}}\tilde{\psi}^{(w_1)\pm}_{-\frac 12}+\sqrt{\frac{w_2}{w_1+w_2}}\tilde{\psi}^{(w_2)\pm}_{-\frac 12}\,,
\\
\tilde{\bar\psi}^{{\rm diag}\pm}_{-\frac 12}=\sqrt{\frac{w_1}{w_1+w_2}}\tilde{\bar\psi}^{(w_1)\pm}_{-\frac 12}+\sqrt{\frac{w_2}{w_1+w_2}}\tilde{\bar\psi}^{(w_2)\pm}_{-\frac 12}\,,
\fe
 The Clifford generators orthogonal to the diagonal ones are
\ie
\tilde{\psi}^{{\rm oth}\pm}_{-\frac 12}=\sqrt{\frac{w_2}{w_1+w_2}}\tilde{\psi}^{(w_1)\pm}_{-\frac 12}-\sqrt{\frac{w_1}{w_1+w_2}}\tilde{\psi}^{(w_2)\pm}_{-\frac 12}\,,
\\
\tilde{\bar\psi}^{{\rm oth}\pm}_{-\frac 12}=\sqrt{\frac{w_2}{w_1+w_2}}\tilde{\bar\psi}^{(w_1)\pm}_{-\frac 12}-\sqrt{\frac{w_1}{w_1+w_2}}\tilde{\bar\psi}^{(w_2)\pm}_{-\frac 12}\,.
\fe
With these, we decompose the space of the product states \eqref{eqn:product_state} into four quartets under the diagonal right-moving Clifford algebra as
\ie\label{eqn:(w1,w2)}
V_{(w_1, w_2)}&=V_{(w_1, w_2){P_1}}\oplus V_{(w_1, w_2){P_2}}\oplus V_{(w_1, w_2){P_3}}\oplus V_{(w_1, w_2){P_4}}\,,
\fe
where the top component subspaces of $V_{(w_1,w_2)P_i}$ are
\ie
V^{\rm top}_{(w_1, w_2){P_1}}&:= \,\textrm{span}
\left(\tilde{\psi}^{{\rm diag}+}_{-\frac 12}\tilde{\bar\psi}^{{\rm diag}+}_{-\frac 12}\ket{\Psi^{w_1,w_2}_{ij}}
\right)\,,
&&
\tilde j = \frac{w_1+w_2}{2}\,,
\\
V^{\rm top}_{(w_1, w_2){P_2}} &:=\,\textrm{span} \left(
\tilde{\psi}^{{\rm diag}+}_{-\frac 12}\tilde{\bar\psi}^{{\rm diag}+}_{-\frac 12}\tilde{\psi}^{{\rm oth}+}_{-\frac 12}\tilde{\bar\psi}^{{\rm oth}+}_{-\frac 12}\ket{\Psi^{w_1,w_2}_{ij}}
\right)\,,
&&
\tilde j = 
\frac{w_1+w_2+2}{2}\,,
\\
V^{\rm top}_{(w_1, w_2){P_3}}&:=\,\textrm{span} \left(
\tilde{\psi}^{{\rm diag}+}_{-\frac 12}\tilde{\bar\psi}^{{\rm diag}+}_{-\frac 12}\tilde{\psi}^{{\rm oth}+}_{-\frac 12}\ket{\Psi^{w_1,w_2}_{ij}}\right)\,,
&&\tilde j = \frac{w_1+w_2+1}{2}\,,
\\
V^{\rm top}_{(w_1, w_2){P_4}}&:=\,\textrm{span} \left(
\tilde{\psi}^{{\rm diag}+}_{-\frac 12}\tilde{\bar\psi}^{{\rm diag}+}_{-\frac 12}\tilde{\bar\psi}^{{\rm oth}+}_{-\frac 12}\ket{\Psi^{w_1,w_2}_{ij}} \right)\,,
&& 
\tilde j = \frac{w_1+w_2+1}{2} \,,
\fe
where $\ket{\Psi^{w_1,w_2}_{ij}}:=\ket{(w_1)_{i,--},(w_2)_{j,--}}$.\footnote{More precisely, we map the states in \eqref{eqn:(w1,w2)} to $S_N$ invariants by summing over their $S_N$ images as in \eqref{eqn:SN_symmetrization}.} We see that the quartets in $V_{(w_1, w_2){P_3}}$ and $V_{(w_1, w_2){P_4}}$ have the same $\tilde j$ as the single-cycle $(w_1+w_2)$ quartets, whose top component subspace is
\ie
V^{\rm top}_{(w_1+w_2)}&=\,\textrm{span} \left(
\tilde{\psi}^{{\rm diag}+}_{-\frac 12}\tilde{\bar\psi}^{{\rm diag}+}_{-\frac 12}\ket{\Psi^{w_1+w_2}_{k}} \right)\,,
&& 
\tilde j = \frac{w_1+w_2+1}{2} \,,
\fe
where $\ket{\Psi^{w_1+w_2}_{k}}:=\ket{(w_1+w_2)_{k,--}}$. By the R-charge conservation, there are no $Q$ and $Q^\dagger$ maps between $V_{(w_1, w_2){P_3}}^{\textrm{top}}$, $V_{(w_1, w_2){P_4}}^{\textrm{top}}$ and $V_{(w_1+w_2)}^{\textrm{top}}$. The possible $Q$ and $Q^\dagger$ actions are
\ie\label{p34_maps}
    \begin{tikzcd}[row sep=2ex, column sep=large, /tikz/column 3/.append style={anchor=base west}]
      & ~ & \hspace{-1.5cm} \bigoplus_{n}\left(V_{(w_1,w_2-n,n)} \oplus V_{(w_1-n,w_2,n)}\right)^{\rm top}_{\tilde j = \frac{w_1+w_2+2}2}
    \\
    V^{\rm top}_{(w_1, w_2){P_{3,4}}} \arrow[ru,"Q_{\rm split}"]  
    \arrow[rd,swap,"(Q^\dagger)_{\rm split}"]
    \\
     & ~ & \hspace{-1.5cm} \bigoplus_{n}\left(V_{(w_1,w_2-n,n)} \oplus V_{(w_1-n,w_2,n)}\right)^{\rm top}_{\tilde j = \frac{w_1+w_2}2}
    \end{tikzcd}
\fe
where $(\cdots)^{\rm top}_{\tilde j=\frac{w_1+w_2}2}$ means projecting onto the top subspace with $\tilde j=\frac{w_1+w_2}2$.

Our goal is to find BPS states in $V^{\rm top}_{(w_1, w_2)P_{3,4}}$.
A first natural ansatz is to consider the {\it composite} two-cycle state
\ie\label{eqn:uvw1w2_P34}
\ket{u_1,u_2}_{w_1+w_2}:=u_1^i u_2^j\ket{\{(w_1)_i,(w_2)_j\}}\in V^{\rm top}_{(w_1, w_2)P_{3,4}}\,,
\fe
where we suppressed the anti-holomorphic indices $\alpha,\beta$, since they have already been specified by the fact that the state is inside $V^{\rm top}_{(w_1, w_2)P_{3,4}}$. $u_1^i$ and $u_2^i$ are coefficients that will be chosen later. Since $Q=Q_{\rm split}$ on the space $V^{\rm top}_{(w_1, w_2)P_{3,4}}$, the only nontrivial $Q$-action is
\ie
Q_{\rm split}\ket{u_1,u_2}_{w_1+w_2} &= \sum_{n=1}^{w_1-1}u_1^i u_2^j c^{(w_1-n)_{k_1},n_{k_2}}_{(w_1)_i}\ket{\{(w_1-n)_{k_1},n_{k_2},(w_2)_j\}}
\\
&\quad+\sum_{n=1}^{w_2-1}u_1^i u_2^j c^{(w_2-n)_{k_1},n_{k_2}}_{(w_2)_j}\ket{\{(w_1)_i,(w_2-n)_{k_1},n_{k_2}\}}\,,
\fe
where the repeated indices are summed. The state \eqref{eqn:uvw1w2_P34} is $Q$-closed, if we choose the vectors $u^i$ and $v^j$ that satisfy
\ie\label{eqn:closedness_conditions}
&u_1^i c^{(w_1-n)_{k_1},n_{k_2}}_{(w_1)_i}= 0\,,\quad u_2^j c^{(w_2-n)_{k_1},n_{k_2}}_{(w_2)_j}= 0\,.
\fe
These conditions are the same conditions for the states 
\ie\label{eqn:single-cycle-cohos}
\ket{u_1}_{w_1}:=u_1^i\ket{\{(w_1)_{i,\pm\pm}\}}\,,\quad \ket{u_2}_{w_2}:=u_2^j\ket{\{(w_2)_{j,\pm\pm}\}}
\fe
being $Q$-closed in the rank $N=w_1$ and $N=w_2$ theories, respectively. A quick way to see that the $Q$-closedness of the single-cycle states \eqref{eqn:single-cycle-cohos} implies the $Q$-closedness of the composite two-cycle state \eqref{eqn:uvw1w2_P34} is to note that the supercharge $Q_{\rm split}$ satisfies the Leibniz rule when acting on multi-cycle states. 

By a similar argument, we know that if the single-cycle states \eqref{eqn:single-cycle-cohos} are $Q^\dagger$-closed then the composite two-cycle state \eqref{eqn:uvw1w2_P34} is also $Q^\dagger$-closed.

Therefore, to guarantee that the state \eqref{eqn:uvw1w2_P34} is BPS, we will assume that the states in \eqref{eqn:single-cycle-cohos} are BPS. Note that the requirement of a single-cycle BPS \emph{state} is stronger than a representative for a single-cycle BPS \emph{cohomology class}; given the latter, the corresponding BPS state under Hodge duality can generally involve the addition of a multi-cycle $Q$-exact term. One may hence worry that our premise is too strong to be useful. Fortunately, there are plenty of examples of single-cycle BPS states. As discussed, the left-moving ${\rm SU}(1,1|2)$ descendants of the single-cycle $\frac12$-BPS states are monotone BPS states. We have also found explicit examples of length-2 single-cycle fortuitous BPS states in \cite{Chang:2025xx}. It is expected that there are more single-cycle fortuitous BPS states with longer lengths.

When are composite two-cycle BPS states fortuitous? A general criterion is not yet known, but we conjecture that they are fortuitous whenever at least one constituent single-cycle state is fortuitous. A detailed analysis will appear in the follow-up work \cite{Chang:2025xx}.

\subsection{Generalization to multi-cycle states}\label{sec_multi_cycle}
Our construction above generalizes to multi-cycles, following by selection rules forbidding the joining processes, for example, for the 3-cycle case, the possible joining processes are
\ie\label{eqn:3cycle-joining}
&(w_1, w_2, w_3) \quad \xlongrightarrow{Q_{\rm join}}
    \quad (w_1+w_2,w_3) \oplus \text{(2 perm)}\,,
\\
&(w_1, w_2, w_3)\quad \xlongrightarrow{(Q^\dagger)_{\rm join}}\quad   (w_1+w_2,w_3) \oplus \text{(2 perm)}\,.
\fe
The selection rule relies on the conservation of both the right-moving R-charge $\tilde j$ and the outer automorphism charge $\widetilde K$, which is preserved under the supercharge $Q$-action, since the action of the deformed supercharge (\ref{ThreePT}) amounts to extracting the OPE between the initial state and a single-cycle length-2 operator with right-moving outer automorphism charge $\tilde{K}=0$.

On the right hand sides of \eqref{eqn:3cycle-joining}, the $(w_1,w_2,w_3)$-sector contains 16 quartets, where the $(\tilde j,\tilde K)$ charges of their top components are listed in Table~\ref{tab:(w1,w2,w3)}. On the left hand sides of \eqref{eqn:3cycle-joining}, the $(\tilde j,\tilde K)$ charges of the top components of the two-cycles are listed in Table~\ref{tab:2cycle_jJ}. Now, it is easy to see that the states in the subspace of the $(w_1,w_2,w_3)$-sector with $\tilde j=\frac{w_1+w_2+w_3+1}2$ and $\widetilde K=\pm 1$ do not participate in the joining processes \eqref{eqn:3cycle-joining}.

\begin{table}[H]
\centering
\begin{tabular}{|c|c|c|}
\hline
$\tilde j$ & $\widetilde K$ & degeneracy \\ \hline
$\frac{w_1+w_2+w_3-1}2$     & $0$ & 1 \\
$\tfrac{w_1+w_2+w_3}{2}$ & $-\tfrac{1}{2},\, \tfrac{1}{2}$ & 2 \\
$\frac{w_1+w_2+w_3+1}2$     & $\{-1,\, 0,\, 1\}$ & \{1,4,1\} \\
$\tfrac{w_1+w_2+w_3+2}{2}$ & $-\tfrac{1}{2},\, \tfrac{1}{2}$ & 2 \\
$\tfrac{w_1+w_2+w_3+3}{2}$     & $0$ & 1\\
\hline
\end{tabular}
\caption{\label{tab:(w1,w2,w3)}The right-moving $R$-charge $\tilde j$ and the outer automorphism charge $\widetilde K$ of the three-cycle $(w_1,w_2,w_3)$ states.}
\end{table}

\begin{table}[H]
\centering
\begin{tabular}{|c|c|c|}
\hline
$(w_1+w_2,w_3)$, $(w_2+w_3,w_1)$, $(w_3+w_1,w_2)$-sectors & $\tilde j$ & $\widetilde K$ \\ \hline\hline
$P_1$ & $\frac{w_1+w_2+w_3}{2}$     & $0$ \\
$P_3$, $P_4$ & $\frac{w_1+w_2+w_3+1}{2}$ & $-\tfrac{1}{2},\, \tfrac{1}{2}$ \\
$P_2$ & $\frac{w_1+w_2+w_3+2}{2}$     & $ 0$ \\
\hline
\end{tabular}
\caption{\label{tab:2cycle_jJ}The right-moving $R$-charge $\tilde j$ and the outer automorphism charge $\widetilde K$ of the two-cycle $(w_1+w_2,w_3)$ states.}
\end{table}

\subsection{Black hole bound states and massive stringy excitations}
\label{sec_two_center_massive}

Depending on whether the constituents are monotone or fortuitous, composite BPS states have different gravitational interpretations.

\paragraph{Fortuitous-fortuitous: Black hole bound states}
It was conjectured \cite{Chang:2024zqi} that fortuitous states are dual to typical black hole microstates, and hence it is natural to interpret $u^i v^j\ket{\{(w_1)_i,(w_2)_j\}}$ as a threshold bound state of two black holes, which geometrically arises in the near horizon limit of two-centered black hole solutions \cite{Giusto:2004id,Giusto:2004ip,deBoer:2008fk} resulting in e.g. nontrivial fibrations of $\mathrm{S}^3$ over AdS.\footnote{Two-centered black holes in 4d lift \cite{Gaiotto:2005gf, Gaiotto:2005xt} to black rings \cite{Elvang:2004rt,Bena:2004de,Elvang:2004ds,Gauntlett:2004qy} in 5d, and one can consider the near horizon limit of the latter. For the microstate counting of black rings, see \cite{Cyrier:2004hj,Halder:2023kza}.}

\paragraph{Fortuitous-monotone: massive stringy excitations}

A state \(u_1^i\ket{\{(w_1)_i\}}\) representing a fortuitous cohomology class at \( N = w_1 \) becomes non-BPS for \( N > w_1 \). In the large \( N \) limit (\( N \gg w_1 \)), we expect it to be dual to a massive stringy excitation in vacuum ${\rm AdS}_3\times {S}^3\times T^4$. 

Now, consider the composites of a fortuitous state \(u_1^i\ket{\{(w_1)_i\}}\) with a monotone state \(u_2^j\ket{\{(w_2)_{j}\}}\), the latter of which is dual to the superstrata geometry \cite{Bena:2015bea, Bena:2016agb, Bena:2017xbt, Bakhshaei:2018vux, Ceplak:2018pws, Heidmann:2019zws, Heidmann:2019xrd, Shigemori:2020yuo}. Hence, in the large \( N \) limit (\( N = w_1 + w_2 \gg w_1 \)), it is natural to propose that the bulk dual of the composite state \(u_1^i u_2^j\ket{\{(w_1)_i,(w_2)_j\}}\) 
is a massive stringy excitation on the superstrata geometry. Remarkably, the non-BPS massive stringy excitation in the ${\rm AdS}_3\times {S}^3\times T^4$ vacuum becomes BPS on the Lunin-Mathur geometry or superstrata background.

\section*{Acknowledgements} 
We would like to thank Luis Apolo, Iosif Bena, Nathan Benjamin, Nathan Brady, Yiming Chen, Matthias Gaberdiel, Bin Guo, Marcel Hughes, Ji Hoon Lee, Samir D. Mathur, Beat Nairz, Masaki Shigemori, Zixia Wei, Ashoke Sen, Ergin Sezgin and Xi Yin for helpful discussions. CC is partly supported by the National Key R\&D Program of China (NO. 2020YFA0713000). CC thanks the hospitality of Southeast University and Peng Huanwu Center for Fundamental Theory, Hefei, where part of the work was done during the visit. HZ thanks the hospitality of Harvard University and Yau Mathematical Sciences Center, where part of the work was done during the visit. The work of HZ is supported in part by DOE Grant No. DE-SC0010813. This research was supported in part by grant NSF PHY-1748958 to the Kavli Institute for Theoretical Physics (KITP).

\appendix

\section{Free field realization of $\mathcal{N}=4$ and covering space calculus}\label{app: coveringspace}

Our convention for the $T^4$ sigma model is that the operator products of the four free complex fields take the form  
\be
\bar \alpha^i(x) \alpha^j(y)\sim \frac{\epsilon^{ij}}{(x-y)^2}, \qquad \bar\psi^{\pm}(x)\psi^{\mp}(y)\sim \pm \frac{1}{x-y}.
\ee
The small $\mathcal{N}=4$ superconformal algebra at $c=6$ is generated by the currents (with normal ordering suppressed)
\ie\label{eqn:TGK_in_alpha-psi}
& T = \bar\alpha^1 \alpha^2-\alpha^1 \bar\alpha^2+ \frac{1}{2}(-\bar\psi^- \partial \psi^++\psi^-\partial\bar\psi^+-\psi^+\partial\bar\psi^-+\bar\psi^+\partial\psi^-), \\
& G^{'+} = \bar\alpha^2\psi^++\alpha^2\bar\psi^+,\qquad  G^{+}=-\bar\alpha^1\psi^+-\alpha^1\bar\psi^+, \\
& G^{'-} = \bar\alpha^1\psi^-+\alpha^1\bar\psi^-,\qquad G^{-}=\bar\alpha^2\psi^-+\alpha^2\bar\psi^-, \\
& J^+ = \bar\psi^+ \psi^+, \qquad J^-=-\bar\psi^-\psi^-, \qquad J^3 = \frac{1}{2} (\bar\psi^+\psi^-+\bar\psi^-\psi^+). 
\fe
There is an additional SU(2) acting as outer-automorphism on the small $\mathcal{N}=4$ algebra.
\be\label{eqn:outer_su2}
K^+ = -\bar\psi^+ \bar\psi^-, \qquad K^-=-\psi^+\psi^-, \qquad K^3 = \frac{1}{2} (\bar\psi^+\psi^--\bar\psi^-\psi^+).
\ee
Bosonizing the fermions proves useful for simplifying the calculation of correlation functions in the covering space
\be
\psi^+(z) = e^{i\varphi(z)},\quad \bar \psi^-(z) = - e^{-i \varphi(z)},\quad \bar\psi^+(z) = e^{i\varphi'(z)}, \quad \psi^-(z) = e^{-i\varphi'(z)}
\ee
Note that all fermion fields anticommute with each other, whereas the bosonized fields commute. To properly account for this distinction, cocycle factors must be introduced, as discussed in \cite{Polchinski:1998rq, Gaberdiel:2024nge}. 

We provide the corresponding representations of single cycle $\frac{1}{2}$-BPS  states in the covering space:
\ie
\ket{w_{\pm\pm,\pm\pm}}\leftrightarrow e^{i\frac{w\pm1}{2} \varphi(z) + i\frac{w\pm1}{2} \varphi'(z) -i \frac{w\pm1}{2}\tilde \varphi(\bar z) -i\frac{w\pm1}{2}\tilde \varphi'(\bar z) }
\fe
The lift of \( V(G_{-\frac{1}{2}}^-\ket{\textrm{BPS}_-}_2)(1) \) to the covering space takes the form:
\be
\bar{\alpha}^1(1) \phi^\dagger_2(1) + \alpha^1(1) \bar{\phi}^\dagger_2(1),
\ee
where \( \phi^\dagger_2(1) \) and \( \bar{\phi}^\dagger_2(1) \) represent the bottom components of the corresponding spin fields. Their bosonizations in the covering space are of form
\be
\phi^\dagger_2(z) = e^{\frac{i}{2}(-\varphi(z)+\varphi'(z)+\tilde{\varphi}(\bar z)+\tilde{\varphi}'(\bar z))}, \quad \bar\phi^\dagger_2(z) = e^{\frac{i}{2}(\varphi(z)-\varphi'(z)+\tilde{\varphi}(\bar z)+\tilde{\varphi}'(\bar z))}
\ee
The following formula is used in the evaluation of fermion correlation functions:
\be
\langle \prod_i e^{i\epsilon_i \varphi(z_i)}\rangle= \prod_{i<j} z_{ij}^{\epsilon_i \epsilon_j},\quad \quad \sum_i \epsilon_i = 0 
\ee

\section{Higher order expansions of partition functions and indices}\label{app:N=2,3indx}

Expanding the index \eqref{eqn:NS_modified_index}, we find the explicit expansion formulae for the $N=2$ and $N=3$ modified indices:
\ie\label{eqn:m_ind_N2}
I^{\mathrm{Sym}^2(T^4)}_\mathrm{NS}=&\frac{2}{q^{\frac12}}+\left(y+\frac{1}{y}\right)-q^{\frac12}\left(6 y^2+12+\frac{6}{y^2}\right)+q \left(y^3+39 y+\frac{39}{y}+\frac{1}{y^3}\right)
\\
&+2 q^{\frac32}
   \left(y^4-28 y^2-96-\frac{28}{y^2}+\frac{1}{y^4}\right)+q^2 \left(39 y^3+513 y+\frac{513}{y}+\frac{39}{y^3}\right)
\\
&-q^{\frac52} \left(12 y^4+708 y^2+2032+\frac{708}{y^2}+\frac{12}{y^4}\right)
\\
&+q^3 \left(y^5+513 y^3+4382 y+\frac{4382}{y}+\frac{513}{y^3}+\frac{1}{y^5}\right)+O\left(q^{\frac72}\right)\,,
\fe
and
\ie\label{eqn:m_ind_N3}
I^{\mathrm{Sym}^3(T^4)}_\mathrm{NS}=&\frac{3}{q^{\frac34}}+q^{\frac14} \left(y^2+8+\frac{1}{y^2}\right)-8 q^{\frac34} \left(y^3+7 y+\frac{7}{y}+\frac{1}{y^3}\right)
\\
&+q^{\frac54} \left(y^4+152
   y^2+513+\frac{152}{y^2}+\frac{1}{y^4}\right)
\\
&-16 q^{\frac74} \left(13 y^3+127 y+\frac{127}{y}+\frac{13}{y^3}\right)
\\
&+q^{\frac94} \left(3 y^6+152 y^4+4382
   y^2+11576+\frac{4382}{y^2}+\frac{152}{y^4}+\frac{3}{y^6}\right)
\\
&-8 q^{\frac{11}4} \left(7 y^5+702 y^3+4511 y+\frac{4511}{y}+\frac{702}{y^3}+\frac{7}{y^5}\right)+O\left(q^{\frac{13}4}\right)\,.
\fe

The short characters $\chi_0$, $\chi_1$ of the $c=12$ contracted large ${\cal N}=4$ superconformal algebra have the expansions 
\ie\label{eqn:N=2_short_char_j=0}
\chi_{0}=&\frac{1}{q^{\frac12}}-2\left(y+\frac{1}{y}\right)+q^{\frac12} \left(2 y^2+\frac{2}{y^2}+9\right)-2 q\left(y^3+9 y+\frac{9}{y}+\frac{1}{y^3}\right)
\\
&+q^{\frac32}
   \left(y^4+22 y^2+60+\frac{22}{y^2}+\frac{1}{y^4}\right)-2 q^2 \left(9 y^3+55 y+\frac{55}{y}+\frac{9}{y^3}\right)
\\
&+q^{\frac52} \left(9 y^4+132
   y^2+305+\frac{132}{y^2}+\frac{9}{y^4}\right)
\\
&-2 q^3 \left(y^5+55 y^3+266 y+\frac{266}{y}+\frac{55}{y^3}+\frac{1}{y^5}\right)+O\left(q^{\frac72}\right)\,,
\fe
and
\ie\label{eqn:N=2_short_char_j=1}
\chi_1=&-\left(y+\frac1y\right)+2 q^{\frac12} \left(y^2+3+\frac{1}{y^2}\right)-q\left(y^3+15 y+\frac{15}{y}+\frac{1}{y^3}\right)
\\
&+q^{\frac32} \left(20 y^2+54+\frac{20}{y^2}\right)-q^2 \left(15 y^3+113 y+\frac{113}{y}+\frac{15}{y^3}\right)
\\
&+q^{\frac 52} \left(6 y^4+144 y^2+338+\frac{144}{y^2}+\frac{6}{y^4}\right)
\\
&-q^3 \left(y^5+113 y^3+630 y+\frac{630}{y}+\frac{113}{y^3}+\frac{1}{y^5}\right)+O\left(q^{7/2}\right)\,.
\fe
The expansion for ${\cal S}_1$ in the $N=2$ BPS partition function \eqref{eqn:BPS_pf} is
\ie
{\cal S}_1=&-5\left(y+\frac{1}{y}\right)+10 q^{\frac12} \left(y^2+3+\frac{1}{y^2}\right)-5 q \left(y^3+15 y+\frac{15}{y}+\frac{1}{y^3}\right)
\\
&+4 q^{\frac32} \left(25
   y^2+78+\frac{25}{y^2}\right)-q^2 \left(75 y^3+733 y+\frac{733}{y}+\frac{75}{y^3}\right)
\\
&+q^{\frac52} \left(30 y^4+972 y^2+2642+\frac{972}{y^2}+\frac{30}{y^4}\right)
\\
&-q^3
   \left(5 y^5+733 y^3+5446 y+\frac{5446}{y}+\frac{733}{y^3}+\frac{5}{y^5}\right)+O\left(q^{\frac72}\right)
\,. \label{QuarterBPSDege_2}
\fe

\section{Cohomology Classes at $h=\frac{3}{2},j=\frac{1}{2}$}\label{sec:Cohomology_h32}
For untwisted $(1,1)_{S_1}$ sector, we have 
\bea
&&(-\bar\psi^{[i]+}_{-\frac{1}{2}}\psi^{[i]-}_{-\frac{1}{2}}\psi^{[i]+}_{-\frac{1}{2}}+\bar\psi^{[j]+}_{-\frac{1}{2}}\psi^{[i]-}_{-\frac{1}{2}}\psi^{[i]+}_{-\frac{1}{2}}+\psi^{[j]+}_{-\frac{1}{2}}\psi^{[i]+}_{-\frac{1}{2}}\bar\psi^{[i]-}_{-\frac{1}{2}} ) \ket{1_{--},1_{--}}\crcr
&& \bar\psi^{[i]+}_{-\frac{1}{2}}\psi^{[i]+}_{-\frac{1}{2}}\bar\psi^{[i]-}_{-\frac{1}{2}}-\bar\psi^{[j]+}_{-\frac{1}{2}}\bar\psi^{[i]+}_{-\frac{1}{2}}\psi^{[i]-}_{-\frac{1}{2}}+ \psi^{[j]+}_{-\frac{1}{2}}\bar\psi^{[i]+}_{-\frac{1}{2}}\bar\psi^{[i]-}_{-\frac{1}{2}}\ket{1_{--},1_{--}}\crcr
&&(\bar\psi^{[j]+}_{-\frac{1}{2}}\psi^{[i]-}_{-\frac{1}{2}}\psi^{[i]+}_{-\frac{1}{2}}+\psi^{[j]+}_{-\frac{1}{2}}\bar\psi^{[i]+}_{-\frac{1}{2}}\psi^{[i]-}_{-\frac{1}{2}})\ket{1_{--},1_{--}}\crcr
&&(\bar\psi^{[i]+}_{-\frac{1}{2}}\psi^{[i]+}_{-\frac{1}{2}}\bar\psi^{[i]-}_{-\frac{1}{2}}-\bar\psi^{[j]+}_{-\frac{1}{2}}\bar\psi^{[i]+}_{-\frac{1}{2}}\psi^{[i]-}_{-\frac{1}{2}}+\bar\psi^{[j]+}_{-\frac{1}{2}}\psi^{[i]+}_{-\frac{1}{2}}\bar\psi^{[i]-}_{-\frac{1}{2}})\ket{1_{--},1_{--}}\crcr
&&(\bar\psi^{[i]+}_{-\frac{1}{2}}\psi^{[i]+}_{-\frac{1}{2}}\bar\psi^{[i]-}_{-\frac{1}{2}}+\bar\psi^{[j]+}_{-\frac{1}{2}}\bar\psi^{[i]+}_{-\frac{1}{2}}\psi^{[i]+}_{-\frac{1}{2}})\ket{1_{--},1_{--}}\crcr
&& (-\bar\psi^{[i]+}_{-\frac{1}{2}}\psi^{[i]-}_{-\frac{1}{2}}\psi^{[i]+}_{-\frac{1}{2}}+\psi^{[j]-}_{-\frac{1}{2}}\bar\psi^{[i]+}_{-\frac{1}{2}}\psi^{[i]+}_{-\frac{1}{2}})\ket{1_{--},1_{--}}\crcr
&&(\psi^{[i]+}_{-\frac{1}{2}}\alpha^{[i]2}_{-1}+\psi^{[j]+}_{-\frac{1}{2}}\alpha^{[i]2}_{-1})\ket{1_{--},1_{--}},\quad (-\bar\psi^{[i]+}_{-\frac{1}{2}}\alpha^{[i]2}_{-1}+\psi^{[j]+}_{-\frac{1}{2}}\bar\alpha^{[i]2}_{-1})\ket{1_{--},1_{--}}\crcr
&&(-\bar\psi^{[i]+}_{-\frac{1}{2}}\alpha^{[i]1}_{-1}+\psi^{[j]+}_{-\frac{1}{2}}\bar\alpha^{[i]1}_{-1})\ket{1_{--},1_{--}},\quad  (\psi^{[i]+}_{-\frac{1}{2}}\alpha^{[i]1}_{-1}+\psi^{[j]+}_{-\frac{1}{2}}\alpha^{[i]1}_{-1})\ket{1_{--},1_{--}}\crcr
&&(\bar\psi^{[i]+}_{-\frac{1}{2}}\alpha^{[i]2}_{-1}+\bar\psi^{[j]+}_{-\frac{1}{2}}\alpha^{[i]2}_{-1})\ket{1_{--},1_{--}},\quad (\bar\psi^{[i]+}_{-\frac{1}{2}}\bar\alpha^{[i]2}_{-1}+\bar\psi^{[j]+}_{-\frac{1}{2}}\bar\alpha^{[i]2}_{-1})\ket{1_{--},1_{--}}\crcr
&& (\bar\psi^{[i]+}_{-\frac{1}{2}}\bar\alpha^{[i]1}_{-1}+\bar\psi^{[j]+}_{-\frac{1}{2}}\bar\alpha^{[i]1}_{-1})\ket{1_{--},1_{--}} ,\quad  (\bar\psi^{[i]+}_{-\frac{1}{2}}\alpha^{[i]1}_{-1}+\bar\psi^{[j]+}_{-\frac{1}{2}}\alpha^{[i]1}_{-1})\ket{1_{--},1_{--}}\crcr
&&(\bar\psi^{[i]+}_{-\frac{1}{2}}\alpha^{[i]2}_{-1}+\psi^{[j]+}_{-\frac{1}{2}}\bar\alpha^{[i]2}_{-1})\ket{1_{--},1_{--}},\quad (\bar\psi^{[i]+}_{-\frac{1}{2}}\alpha^{[i]1}_{-1}+\psi^{[j]+}_{-\frac{1}{2}}\bar\alpha^{[i]1}_{-1})\ket{1_{--},1_{--}}\crcr
&& \psi^{[i]+}_{-\frac{3}{2}}\ket{1_{--},1_{--}},\quad \bar\psi^{[i]+}_{-\frac{3}{2}}\ket{1_{--},1_{--}}
\eea
where summation of i, j ranging from 1 to 2 is implicit to ensure orbifold invariance.
For twisted sector, we have 
\bea
&& \bar\psi^+_{-\frac{1}{2}}\psi^-_0\psi^+_{-\frac{1}{2}}\bar\psi^-\ket{2_{--}},\quad (-\psi^-_0\psi^+_{_{-\frac{1}{2}}}\bar\alpha^2_{-\frac{1}{2}}+\psi^+_{-\frac{1}{2}}\bar\psi^-_0\alpha^2_{-\frac{1}{2}})\ket{2_{--}},\crcr
&&\psi^+_{-\frac{1}{2}}\bar\psi^-_0\alpha^1_{-\frac{1}{2}}\ket{2_{--}},\quad \bar\psi^+_{-\frac{1}{2}}\psi^-\bar\alpha^2_{-\frac{1}{2}}+ \bar\psi^+_{-\frac{1}{2}}\bar\psi^-_0\alpha^2_{-\frac{1}{2}}\ket{2_{--}}\crcr
&& \bar\psi^+_{-\frac{1}{2}}\bar\psi^-_0 \alpha^1_{-\frac{1}{2}}\ket{2_{--}},\quad \psi^+_{-\frac{1}{2}}\bar\psi^-_{-\frac{1}{2}},\quad (2 \alpha^1_{-\frac{1}{2}}\bar\alpha^2_{-\frac{1}{2}}+\psi^+_{-1}\bar\psi^-_0) \ket{2_{--}},\quad \psi^-_{-\frac{1}{2}}\psi^+_{-\frac{1}{2}}\ket{2_{--}}\crcr
&&\alpha^2_{-1}\ket{2_{--}},\quad \alpha^1_{-1}\ket{2_{--}},\quad \bar\alpha^1_{-1}\ket{2_{--}},\quad \bar\alpha^2_{-1}\ket{2_{--}}
\eea

The complete list of states in the $(1,1)_A$ sector is too extensive to present here; the essential point is to identify all possible antisymmetric left-moving excitations.

\bibliography{refs}

\bibliographystyle{utphys}

\end{document}